\documentclass[preprint,showpacs,preprintnumbers,amsmath,amssymb,prb]{revtex4}

\usepackage{graphicx}
\usepackage{dcolumn}
\usepackage{bm}

\begin{document}

\preprint{quant-ph/0405069}

\title{On the principles of quantum mechanics}

\author{Eijiro Sakai}
 \email{esakai@sci.u-ryukyu.ac.jp}
\affiliation{
Department of Physics and Earth Sciences, University of the Ryukyus,
Okinawa, 903-0213, Japan
}%

\date{\today}

\begin{abstract}
We propose six principles as the fundamental principles of quantum mechanics: principle of space and time, Galilean principle of relativity, Hamilton's principle, wave principle, probability principle, and principle of indestructibility and increatiblity of particles. 
We deductively develop the formalism of quantum mechanics on the basis of them: we determine the form of the Lagrangian that satisfies requirements of these principles, and obtain the Schr\"odinger equation from the Lagrangian.
We also derive the canonical commutation relations.
Then we adopt the following four guide lines. 
First, we do not premise the relations between dynamical variables in classical mechanics. 
Second, we define energy, momentum, and angular momentum as the constants of motion that are derived from homogeneity and isotropy in space and time on the basis of principle of space and time. 
Since energy and momentum are quantitatively defined in classical mechanics, we define them in quantum mechanics so that the corresponding conservation laws are satisfied in a coupling system of a quantum particle and a classical particle. 
Third, we define Planck's constant and the mass of a particle as proportionality constants between energy and frequency due to one of Einstein-de Broglie formulas and between momentum and velocity, respectively.
We shall obtain the canonical commutation relations and the Schr\"odinger equation for a particle in an external field in the definitive form. 
We shall also prove that relations between dynamical variables in quantum mechanics have the same forms for those in classical mechanics.

\end{abstract}

\maketitle


\section{\label{sec1}INTRODUCTION}

About 80 years were passed after quantum mechanics was established. 
Almost all physicists recognize that quantum mechanics is completely right and 
that it provides the basis for the modern physics. Quantum mechanics, however, has the flaw that it does not form a self-contained formalism.

In most treatises on quantum mechanics, to begin with, the Schr\"odinger 
equation is introduced heuristically with the help of Einstein-de Broglie 
formulas, and next, ``interpretations'' are ad hoc added to which are required to understand results obtained by applying it to specific concrete problems. 
Such developments are useful in order to make readers learn quantum mechanics quickly. 
It is, however, very unsatisfactory since quantum mechanics is the basis of the modern physics.
On the one hand, classical mechanics is deductively established on the basis 
of three laws of motion. 

In the Course of Theoretical Physics, Landau first determine the Lagrangians for the nonrelativistic\footnote{See L.D. Landau and E.M. Lifshitz, ``\textit{Mechanics}'' (Course of Theoretical Physics, Vol. 1) 3rd ed. (Pergamon Press, New York, 1976), Chap.1.} and relativistic\footnote{See L.D. Landau and E.M. Lifshitz, ``\textit{The Classical Theory of Field}'' (Course of Theoretical Physics, Vol. 2) 4th ed. (Pergamon Press, New York, 1975), Chap.2.} particles and the Lagrangian density for the electromagnetic field in vacuum\footnote{See L.D. Landau and E.M. Lifshitz, ``\textit{The Classical Theory of Field}'' (Course of Theoretical Physics, Vol. 2) 4th ed. (Pergamon Press, New York, 1975), Chap.4.} from physical principles and then derive the equations of motion and the Maxwell's equations.
He, however, renounce deriving the Schr\"odinger equation in the same way, because he had believed that it is in priciple impossible to formulate the basic concepts of quantum mechanics without using classical mechanics.\footnote{See L.D. Landau and E.M. Lifshitz, ``\textit{Quantum Mechanics}'' (Course of Theoretical Physics, Vol. 3) 3rd ed. (Pergamon Press, New York, 1975), Sec.1.}

Quantum mechanics includes classical mechanics in the limit; classical mechanics is the approximated theory of quantum mechanics, 
because quantum mechanics is reduced to classical mechanics in the limit in which Planck's constant approaches zero. 
Therefore, the author convinces himself that quantum mechanics should be established on its own proper basis independently 
of classical mechanics.

In order to set quantum mechanics as the foundation of the modern physics in the true sense, 
quantum mechanics must be reconstructed from the ``quantized'' 
mechanics into the true ``quantum'' mechanics. 
The true ``quantum'' mechanics must not be premised on classical mechanics that is approximate to quantum mechanics. 
In the present paper we shall propose six principles as the fundamental principles of 
the true quantum mechanics: principle of space and time, Galilean principle of relativity, Hamilton's principle, wave principle, probability principle, and principle of indestructibility and increatiblity of particles.
We shall deductively develop the formalism of quantum mechanics on the basis of them: we determine the form of the Lagrangian that satisfies requirements of these principles, and obtain the Schr\"odinger equation from the Lagrangian. 
We also derive the canonical commutation relations.
Then we adopt the following four guide lines. 
First, we do not premise the relations between dynamical variables in classical mechanics such as $E = {\bf{p}}^2 /2m$ and ${\bf{L}} = {\bf{r}} \bm{\times} {\bf{p}}$. 
Second, we define energy, momentum, and angular momentum as the constants of motion that are derived from homogeneity and isotropy in space and time on the basis of principle of space and time. 
Since energy and momentum are quantitatively defined in classical mechanics, we define them in quantum mechanics so that the corresponding conservation laws are satisfied in a coupling system of a quantum particle and a classical particle. 
Third, we define Planck's constant and the mass of a partilce as proportionality constants between energy and frequency due to one of Einstein-de Broglie formulas and between momentum and velocity, respectively.
In the interests of simplicity, we assume that the particle has no internal degrees of freedom.

In Sec.~II, we shall propose the six principles of quantum mechanics: principle of space and time, Galilean principle of relativity, Hamilton's principle, wave principle, and probability principle, and principle of indestructibility and increatiblity of particles. 
In Sec.~III, we shall describe basic concepts that are directly derived from the principles: the expectation values of operator, properties of the position operator, and the position probability density of a particle. 
In the following three sections we shall discuss the properties of generators of three fundamental transformations: translation, rotation and Galilean transformation. 
In Sec.~VII, we shall obtain the form of the Lagrangian for a free particle, using the principle of space and time, probability principle, Galilean principle of relativity, and principle of indestructibility and increatibility of particles. The resultant Lagrangian gives the essentially true form of the Schr\"odinger equation for a free particle except for that it includes three undetermined constants.
In Sec.~VIII, we shall show that two of the undetermined constants are Planck's constant and the mass of a particle on the basis of conservation laws of energy and momentum.
In Sec.~IX, we shall obtain the Lagrangian for two interacting particles, and from it we shall find the Lagrangian for a particle in an external field as that for the relative motion. 
Furthermore we shall determine the last undetermined constant by the conservation law of momentum for a system consisting of a quantum particle and a classical particle.
This results that we obtain the general form of the Schr\"odinger equation for a particle in an external field and verify the commutation relations between the position and momentum operators.
In the last section we shall give conclusion. 

We are interested in the logical structure of quantum mechanics, so we shall not pursue mathematical rigorousness.  
We will not dwell on the theory of measurement.

\section{\label{sec2}PRINCIPLES OF QUANTUM MECHANICS}

In this section, we shall propose the six fundamental principles of 
nonrelativistic quantum mechanics, which deals with a microscopic particle 
that is regarded as a particle from the point of view of the classical 
theory. We shall not consider the quantum theory of such as the electromagnetic 
field that is regarded classically as wave. 

\vspace{12pt}

{\bf 1. Principle of Space and Time}

Principle of space and time is stated as
\begin{quote}
\textit{Space is homogeneous, isotropic and flat, which means that space is Euclidean. Time is homogeneous.}
\end{quote}

This is the geometric principle with regard to space and time in which physical phenomena occur. In classical mechanics, this principle is 
implicitly included in the first law of motion, ``the law of inertia''.

\vspace{12pt}

{\bf 2. Galilean Principle of Relativity}

Galilean principle of relativity is stated as
\begin{quote}
\textit{The laws of physics are covariant under Galilean transformations.}
\end{quote}

Since our purpose is to establish nonrelativistic quantum mechanics, 
we should naturally adopt Galilean principle of relativity.

\vspace{12pt}

{\bf 3. Hamilton's Principle}

Hamilton's principle is stated as
\begin{quote}
\textit{The system makes a motion so that its action has the minimum value. The action is defined as the integral over time of the Lagrangian. 
The Lagrangian is a real additive scalar function of variables that represent 
states of a system of interest.
To gurantee the covariance of physical laws, the Lagrangian should be invariant with respect to symmetrical transformations that come from the first two principles. }
\end{quote}

From this principle we obtain equation of motion. 

\vspace{12pt}

{\bf 4. Wave Principle}

Wave principle is stated as
\begin{quote}
\hspace{10 pt} \textit{States of a microscopic particle are represented by state vectors that satisfy the principle of superposition. Especially each state of a microscopic particle is completely determined except for internal degrees of freedom if the position of the particle is specified. All the state vectors, therefore, that represent a particle located at a particular position constitute a complete set. Any dynamical variable is represented by an operator which operates on state vectors.}
\end{quote}

A state vector is usually expressed as $\vert\psi\rangle$, and its adjoint 
$(\vert\psi\rangle)^{\dag}$ is expressed as $\langle\psi\vert$. 
When we need to distinguish an original vector and its adjoint vector, 
we call an original vector a ket and its adjoint a bra after Dirac.\footnote{See P.A.M. Dirac, ``\textit{The Principles of Quantum Mechanics}'' 4th ed. (Oxford University Press, Oxford, 1958), Sec.6.} In the present paper we assume that state vectors are always normalized.
The principle of superposition means that a superposition 
(a linear combination) of two state vectors, each of which represents a 
possible microscopic state, represents also another possible microscopic state.
Dynamical variable operators must be linear in order to satisfy the principle 
of superposition. The properties of them are derived from the 
principles of quantum mechanics, as shall be shown below. The operator 
corresponding to a dynamical variable $A$ is expressed by $\hat{A}$. 

In the interests of simplicity, we assume that the particle has no internal degrees of freedom.

\vspace{12pt}

{\bf 5. Probability Principle}

Probability principle is stated as
\begin{quote}
\hspace{10 pt} \textit{If a state of a particle is given by a normalized state vector 
$\vert\psi\rangle$ that is a superposition of $\vert k \rangle$:
\begin{equation*}
\vert \psi \rangle
=
\sum_{k} c_{k} \vert k \rangle,
\end{equation*}
where $\vert k \rangle$ is the eigenvector corresponding to an eigenvalue 
$a_{k}$ of an operator $\hat{A}$, the result of measurement of the dynamical variable $A$ 
is equal to one of eigenvalues of $\hat{A}$, $a_{k}$, and the probability 
that we obtain $a_{k}$ is given by $\vert c_{k}\vert^{2}$.}
\end{quote}

The probability principle is in close connection with the theory of 
measurement. Since there exists no theory of measurement that is generally 
accepted, we have inaccurately stated the principle here. For example, we have not referred to the ``reduction of wave packet''. We shall revise 
the principle, when the theory of measurement have been established.

The complete theory of measurement, however, is not needed to develop the formalism of quantum mechanics on the basis of principles. 
For it only three theorems are needed. 
First of them is that dynamical variable operators are observable. 
Second is that the expectation value of measurement on a dynamical variable 
can be calculated from the probability principle. The last is that the 
square of the magnitude of the wave function gives the probability 
density of existence of the particle. These three theorems will be derived 
in the next section.

\vspace{12pt}

{\bf 6. Principle of Indestructibility and Increatibility of particles}

Indestructibility and Increatibility is stated as
\begin{quote}
\textit{Particles that quantum mechanics deals with are indestructible and increatible.}
\end{quote}

We require that particles are indestructible and increatible, since quantum mechanics is devoted to low-energy phenomena. This principle is less general than the other principles and auxiliary.

\vspace{12pt}

We shall deductively develop the formalism of quantum mechanics on the basis of these principles in the following sections.

\section{\label{sec3}FUNDAMENTAL CONCEPTS OF QUNATUM MECHANICS}

In this section, we shall describe important concepts directly derived 
from the principles of quantum mechanics.

\subsection{\label{sec31}Property of operators}

From the probability principle, we shall derive the property of the operator corresponding to a dynamical variable. 

The result of measurement of a dynamical variable $A$ must always be a real number.
 Therefore the operator, $\hat{A}$, corresponding to $A$ must be self-adjoint.
Whenever we measure $A$, we should get an eigenvalue of $\hat{A}$ as a result. This requires that the eigenvectors $\vert k \rangle$ of $\hat{A}$ must constitute a complete set:
\begin{equation*}
\sum_{k}\vert k \rangle\langle k \vert=1.
\end{equation*}
Therefore $\hat{A}$ is an observable.

Assume that a normalized state vector $\vert\psi\rangle$ is expanded as
\begin{equation*}
\vert\psi\rangle=\sum_{k}c_{k}\vert k \rangle
\end{equation*}
in terms of eigenvectors $\vert k \rangle$ of a dynamical variable operator$\hat{A}$, 
where the expansion coefficients are given by 
\begin{equation*}
c_{k}=\langle k \vert\psi\rangle.
\end{equation*}
From the probability principle, if the state of the system is given by this state vector, 
the expectation value $\langle\hat{A}\rangle$ is expressed as
\begin{equation*}
\langle\hat{A}\rangle=\sum_{k}\vert c_{k} \vert^{2}a_{k},
\end{equation*}
where $a_{k}$ is the eigenvalue corresponding to the vector $\vert k \rangle$.
This expectation value can be expressed as
\begin{equation*}
\langle\hat{A}\rangle=\sum_{k}\vert c_{k} \vert^{2}a_{k}=
\sum_{k}\langle\psi\vert k \rangle a_{k}\langle k \vert\psi\rangle=
\sum_{k}\langle\psi\vert\hat{A}\vert k \rangle\langle k \vert\psi\rangle=
\langle\psi\vert\hat{A}\vert\psi\rangle,
\end{equation*}
where use has been made of the completeness of the eigenvectors 
$\vert k \rangle$.

\subsection{\label{sec34}Position operator}

We express the position operator of a microscopic particle as 
$\hat{{\bf r}}$ and its eigenvector as $\vert{\bf r}\rangle$;
\begin{equation}
\hat{{\bf r}} \vert{\bf r} \rangle
=
{\bf r} \vert{\bf r} \rangle.
\label{eqPos}
\end{equation}
We assume that eigenvectors of the position operator are not degenerate, because we only deal with particles without internal degrees of freedom in the present paper. 
This follows that
\begin{equation*}
\langle {\bf r} \vert {\bf r}^{\prime} \rangle
=
0 \ \ \ \ \ 
\mathrm{for} 
\ \ \ \ \  
{\bf r} \neq {\bf r}^{\prime}.
\end{equation*}

From the wave principle, all the vectors $\vert{\bf r}\rangle$ should constitute a complete set, so that the equation
\begin{equation*}
\int d{\bf r}\vert{\bf r}\rangle\langle{\bf r}\vert
=1
\end{equation*}
is valid. The relation
\begin{equation*}
\left(\int d{\bf r}\vert{\bf r}\rangle\langle{\bf r}\vert
\right)^{2}=\int d{\bf r}\int d{\bf r}^{\prime}
\vert{\bf r}\rangle\langle{\bf r}\vert{\bf r}^{\prime}
\rangle\langle{\bf r}^{\prime}\vert=1
\end{equation*}
gives
\begin{equation*}
\langle {\bf r} \vert {\bf r}^{\prime} \rangle
=
\delta ({\bf r}-{\bf r}^{\prime}),
\end{equation*}
which is the orthonormal condition for the eigenvectors of the position 
operator.

From Eq.~(\ref{eqPos}) we obtain the relation
\begin{equation*}
\hat x_j \hat x_k \left| {\bf{r}} \right\rangle  
=
x_j  x_k \left| {\bf{r}} \right\rangle  
=
 \hat x_k \hat x_j \left| {\bf{r}} \right\rangle ,
\end{equation*}
where $\hat x_j $ denotes the $j$-th Cartesian component of ${\bf{\hat r}}$.
The above equation leads that
\begin{equation}
\left[ {\hat x_j ,\hat x_k } \right] = 0,
\label{eqccr1}
\end{equation}
since $\vert{\bf r}\rangle$ constitute a complete set.

\subsection{\label{sec35}Position probability density of the particle}

In accordance with the general expression for the expectation value, 
the expectation value of position is given by
\begin{eqnarray*}
\langle\hat{{\bf r}}\rangle 
&=& 
\langle\psi\vert\hat{{\bf r}}\vert \psi\rangle
=
\int d{\bf r}\langle\psi\vert\hat{{\bf r}}\vert
{\bf r}\rangle\langle{\bf r}\vert\psi\rangle 
\nonumber \\
&=& \int d{\bf r}\langle\psi\vert{\bf r}\rangle{\bf r}\langle
{\bf r}\vert\psi\rangle=\int d{\bf r}~{\bf r}\vert\langle
{\bf r}\vert\psi\rangle\vert^{2},
\end{eqnarray*}
which means that $\vert\langle{\bf r}\vert\psi\rangle\vert^{2}$ 
represents the position probability density of the particle. 
The quantity $\langle{\bf r}\vert\psi\rangle$ is the wave function, 
and is usually designated by $\psi({\bf r})$:
\begin{equation*}
\psi({\bf r})=\langle{\bf r}\vert\psi\rangle
\end{equation*}

Since principle of indestructibility and increatibility of particles, 
the probability finding the particle somewhere is equal to unity independently of time. 
Thus the equation of continuity
\begin{equation}
\displaystyle
\frac{\partial}{\partial t}\vert\psi({\bf r})\vert^{2}+
\mbox{div}{\bf J}=0
\label{eq15}
\end{equation}
must be valid, where {\bf J} is called the probability current density 
and is a real vector. We require that this equation of continuity derives directly from the equation of motion for the state vector. Thus {\bf J} should be expressed with $\psi({\bf r})$ and $\psi^{\ast}({\bf r})$, since the wave function should satisfy a linear equation of motion from the wave principle.

\section{\label{sec4}TRANSLATION}

In this and the next two sections we shall discuss the basic theories of 
translation, rotation and the Galilean transformation, to prepare 
for deriving the equation of motion for the state vector.

\subsection{\label{sec41}Translation and its generator}

The translation operator $\hat{U}({\bf a})$ is defined by
\begin{equation}
\hat{U}({\bf a})\vert{\bf r}\rangle=
\vert{\bf r}+{\bf a}\rangle.
\label{DefU}
\end{equation}
The translation operator must evidently be unitary. From principle of space and time, all translations commute with each other and their translation vectors are additive. Translation operators therefore satisfy the following relation
\begin{equation}
\hat{U}({\bf a})\hat{U}({\bf b})=
\hat{U}({\bf b})\hat{U}({\bf a})=
\hat{U}({\bf a}+{\bf b}).
\label{eq21}
\end{equation}

From the definition of the translation operator we obtain the expression
\begin{equation*}
\hat{U}^{\dag}({\bf a})\hat{{\bf r}}\hat{U}({\bf a})
\vert{\bf r}\rangle=
({\bf r}+{\bf a})\vert{\bf r}\rangle,
\end{equation*}
which means from the completeness of the kets $\vert{\bf r}\rangle$ 
that
\begin{equation}
\hat{U}^{\dag}({\bf a})\hat{{\bf r}}\hat{U}({\bf a})=
\hat{{\bf r}}+{\bf a}.
\label{eq23}
\end{equation}

From Eq.~(\ref{eq21}), 
the translation operator is given by
\begin{equation}
\hat{U}({\bf a})=\exp
\left[
-i\displaystyle\frac{{\bf a}\cdot\hat{{\bf p}}}{\theta}
\right],
\label{eq28}
\end{equation}
where $\hat{{\bf p}}$ is a self-adjoint operator, which is called 
the \textit{translation generator} here after, and must satisfy the commutation relations
\begin{equation}
\left[\hat{p}_{j},\hat{p}_{k}\right]=0.
\label{eqccr2}
\end{equation}
The quantity $\theta$ is a real constant.
Usually, the translation generator is self-evidently identified with momentum operator and $\theta$ is set to be equal to Planck's constant divided by $2 \pi$. 
We do not regard, however, them as self-evident. We shall verify them in Secs.~\ref{sec8} and \ref{Int}, respectively.

From Eqs.~(\ref{eq23}) and (\ref{eq28}), we obtain the commutation relations between the position operator and the translation generator
\begin{equation}
\left[\hat{x}_{j},\hat{p}_{k}\right]=i\theta\delta_{jk}.
\label{eqccr3}
\end{equation}
In similar fashion we can obtain the commutation relation
\begin{equation}
\left[ {{\bf{\hat p}},f\left( {{\bf{\hat r}}} \right)} \right] =  - i\theta \left. {\nabla f\left( {\bf{r}} \right)} \right|_{{\bf{r}} = {\bf{\hat r}}} .
\label{eqccrp}
\end{equation}

\subsection{\label{sec42}Properties of the translation generator and 
its eigenvectors}

From Eqs.~(\ref{DefU}) and (\ref{eq28}), we obtain
\begin{equation}
\hat{{\bf p}}\vert{\bf r}\rangle
=i\theta\nabla\vert{\bf r}\rangle
\label{eq34a}
\end{equation}
and
\begin{equation}
\langle {\bf r} \vert \hat{{\bf p}}
=
-i\theta \nabla \langle {\bf r} \vert.
\label{eqPosR}
\end{equation}
We denote eigenkets of the translation generator $\hat{{\bf p}}$ 
as $\vert{\bf p}\rangle$:
\begin{equation*}
\hat{{\bf p}}\vert{\bf p}\rangle
=
{\bf p}\vert{\bf p}\rangle
\end{equation*}
where {\bf p} is a real vector. Multiplying the above equation by an eigenbra of the position operator 
$\langle{\bf r}\vert$ from the left, we obtain, 
with the help of Eq.~(\ref{eqPosR}), a differential equation
\begin{equation*}
-i\theta\nabla\langle{\bf r}\vert{\bf p}\rangle=
{\bf p}\langle{\bf r}\vert{\bf p}\rangle.
\end{equation*}
We can easily solve this equation to obtain
\begin{equation*}
\langle {\bf r} \vert {\bf p} \rangle
=
Ce^{i{\bf p} \cdot {\bf r} / \theta}.
\end{equation*}

We therefore obtain
\begin{equation*}
\int d{\bf p}\langle{\bf r}\vert{\bf p}\rangle
\langle{\bf p}\vert{\bf r}^{\prime}\rangle=
\vert C \vert^{2}\int d{\bf p}
e^{i{\bf p}\cdot({\bf r}-{\bf r}^{\prime})/\theta}
=(2\pi\theta)^{3}\vert C \vert^{2}
\delta({\bf r}-{\bf r}^{\prime}).
\end{equation*}
It follows from this that
\begin{equation}
\int d{\bf p}
\vert{\bf p}\rangle\langle{\bf p}\vert=1
\label{eq41}
\end{equation}
if we take the normalization constant as
\begin{equation*}
\displaystyle
C=\frac{1}{(2\pi\theta)^{3/2}}.
\end{equation*}
The equation (\ref{eq41}) means that all the eigenvectors for the translation 
generator constitute a complete set. Thus the normalized eigenfunctions of 
the translation generator are
\[
\langle{\bf r}\vert{\bf p}\rangle=\frac{1}{(2\pi\theta)^{3/2}}
e^{i{\bf p}\cdot{\bf r}/\theta}.
\]

The above discussion shows that the translation generator is an observable, 
which implies that the translation generator can correspond to a certain dynamical 
variable.

\subsection{\label{sec43}General property of operators corresponding to 
translational-invariant dynamical variables}

Now we consider a dynamical variable $\hat A$ that is invariant with respect to space 
translation:
\begin{equation*}
\hat{U}^{\dag}({\bf a})\hat{A}\hat{U}({\bf a})=\hat{A}
\end{equation*}
By differentiating both sides of the above equation with respect to {\bf a} and 
putting ${\bf a}=0$, we obtain
\begin{equation*}
\left[\hat{A},\hat{{\bf p}}\right]=0.
\end{equation*}
This means that any translational-invariant dynamical variable commutes with 
the translation generator. The operator $\hat{A}$ is in general represented as
\begin{equation*}
\hat{A}=\int d{\bf p}\int d{\bf p}^{\prime}
\vert{\bf p}\rangle\langle{\bf p}\vert\hat{A}
\vert{\bf p}^{\prime}\rangle\langle{\bf p}^{\prime}\vert=
\int d{\bf p}\int d{\bf p}^{\prime}
\vert{\bf p}\rangle A({\bf p},{\bf p}^{\prime})
\langle{\bf p}^{\prime}\vert
\end{equation*}
in terms of the eigenvectors of the translation generator $\hat{{\bf p}}$ that constitute a complete orthonormal set. 
The off-diagonal elements of $A({\bf p},{\bf p}^{\prime})$ 
must vanish since $\hat{A}$ commutes with $\hat{{\bf p}}$. Thus,
\begin{equation*}
\hat{A}=\int d{\bf p}
\vert{\bf p}\rangle A({\bf p})\langle{\bf p}\vert
=\int d{\bf p}A(\hat{{\bf p}})\vert{\bf p}\rangle
\langle{\bf p}\vert
=A(\hat{{\bf p}}),
\end{equation*}
which shows that $\hat{A}$ is a function only of the translation generator. 
Therefore any translational-invariant dynamical variable is in general 
represented by a function only of the translation generator.

Similarly, a dynamical variable that commutes with the position operator 
is represented by a function only of the position operator; if
\begin{equation*}
\left[\hat{A},\hat{{\bf r}}\right]=0,
\end{equation*}
then
\begin{equation*}
\hat{A}=A(\hat{{\bf r}}).
\end{equation*}

\section{\label{sec5}ROTATION}

\subsection{\label{sec51}Rotation and its generator}

Suppose that a position of a particle, {\bf r}, 
is transformed into ${\bf r}^{\prime}$ by a rotation ${\bm {\varphi}}$
\begin{equation*}
{\bf r}\rightarrow{\bf r}^{\prime}=R({\bm {\varphi}}){\bf r},
\end{equation*}
where $R({\bm {\varphi}})$ is a $3\times3$ orthogonal matrix corresponding to the rotation ${\bm {\varphi}}$. The rotation operator $\hat{V}({\bm {\varphi}})$ is defined by
\begin{equation*}
\hat{V}({\bm {\varphi}})\vert{\bf r}\rangle=
\vert R({\bm {\varphi}}){\bf r}\rangle.
\end{equation*}
The rotation operator must be unitary. From the definition of the rotation operator 
we obtain
\begin{equation}
\hat{V}^{\dag}({\bm {\varphi}})\hat{{\bf r}}\hat{V}({\bm {\varphi}})
=
R({\bm {\varphi}})\hat{{\bf r}}
\label{eq53}
\end{equation}
with the help of completeness of all the state vectors 
$\vert{\bf r}\rangle$.

It is evident that the relation
\begin{equation*}
\hat{V}^{n}({\bm {\varphi}})=\hat{V}(n{\bm {\varphi}})
\end{equation*}
is valid. 
We can therefore represent the rotation operator $\hat{V}({\bm {\varphi}})$, 
which is unitary, as
\begin{equation}
\hat{V}({\bm {\varphi}})=
\exp\left[-\frac{i}{\theta}{\bm {\varphi}}\cdot\hat{{\bf L}}\right],
\label{eq58}
\end{equation}
where the constant $\theta$ has been set the same as that in Eq.~(\ref{eq28}), 
and $\hat{{\bf L}}$ is a self-adjoint operator, which we call 
the \textit{rotation generator}. 
In Sec.~\ref{sec8} we shall verify that the rotation generator is identified with angular momentum operator.
Usually the rotation generator is self-evidently identified with angular momentum operator, but we do not regard it as self-evident and shall verify it in Sec.~\ref{sec8}.

Position {\bf r} is transformed by an infinitesimal rotation 
$\delta{\bm {\varphi}}$ as
\begin{equation*}
R(\delta{\bm {\varphi}}){\bf r}\approx{\bf r}-
{\bf r} \bm{\times} \delta{\bm {\varphi}},
\end{equation*}
which follows, with the help of Eqs.~(\ref{eq53}) and (\ref{eq58}), that
\begin{equation}
\left[ \hat{L}_{j}, \hat{x}_{k} \right]
=
i \theta \varepsilon_{jkl} \hat{x}_{l},
\label{eq61}
\end{equation}
where $ \varepsilon_{jkl} $ is the fully antisymmetric tensor with $ \varepsilon_{123} = 1 $ and if the same letter appears twice as a subscript, the summation over that subscript is meant.

\subsection{\label{sec52}Vector and scalar operators}

We consider how the translation generator is transformed by rotation. 
It follows from the definition of the rotation and translation operators that
\begin{equation*}
\hat{V}^{\dag}({\bm {\varphi}})\hat{U}({\bf a})\hat{V}({\bm {\varphi}})
\vert{\bf r}\rangle
=
\vert{\bf r}+R^{-1}({\bm {\varphi}}){\bf a}\rangle
=
\hat{U}(R^{-1}({\bm {\varphi}}){\bf a})\vert{\bf r}\rangle.
\end{equation*}
This gives
\begin{equation*}
\hat{V}^{\dag}({\bm {\varphi}})\hat{U}({\bf a})\hat{V}({\bm {\varphi}})
=
\hat{U}(R^{-1}({\bm {\varphi}}){\bf a})
\end{equation*}
or
\begin{equation*}
\hat{V}^{\dag}({\bm {\varphi}})\exp\left[-\frac{i}{\theta}
\hat{{\bf p}}\cdot{\bf a}\right]\hat{V}({\bm {\varphi}})
=
\exp\left[-\frac{i}{\theta}\hat{{\bf p}}
\cdot(R^{-1}{\bf a})\right]=\exp\left[-\frac{i}{\theta}
(R\hat{{\bf p}})\cdot{\bf a}\right],
\end{equation*}
which shows that the translation generator is transformed by the rotation as
\begin{equation}
\hat{V}^{\dag}({\bm {\varphi}})\hat{{\bf p}}\hat{V}({\bm {\varphi}})
=
R({\bm {\varphi}})\hat{{\bf p}},
\label{eq65}
\end{equation}
in the same way as the position operator(see Eq.~(\ref{eq53})). This means that the translation 
generator be a vector operator, since a quantity composed of three components 
that is transformed in the same way as the position operator by rotation is 
in general called a vector. 
Equation (\ref{eq65}) yields the commutation relations
\begin{equation}
\left[\hat{L}_{j},\hat{p}_{k}\right]=i\theta\varepsilon_{jkl}\hat{p}_{l}.
\label{eq66}
\end{equation}
Any vector operator $\hat{{\bf A}}$ must also satisfy the commutation relations
\begin{equation}
\left[\hat{L}_{j},\hat{A}_{k}\right]=i\theta\varepsilon_{jkl}\hat{A}_{l}.
\label{eqCRLV}
\end{equation}

Since a quantity that is unchanged under rotation is in general called a 
scalar, a scalar operator should commute with the rotation generator. 
It is easily seen that the inner product of two arbitrary vector operators 
$\hat{{\bf A}}$ and $\hat{{\bf B}}$ commutes with the rotation 
generator
\begin{equation*}
\left[
\hat{{\bf L}},\hat{{\bf A}}\cdot\hat{{\bf B}}
\right]=0,
\end{equation*}
which means that the inner product is a scalar, as should be.

\subsection{\label{sec53}Explicit expression for the rotation generator}

It follows from the commutation relations (\ref{eqccr1}),(\ref{eqccr2}) and (\ref{eqccr3}) that the rotation generator satisfies the commutation relations (\ref{eq61}) and (\ref{eq66}) if
\begin{equation}
\hat{L}_{j}=\varepsilon_{jkl}\hat{x}_{k}\hat{p}_{l}.
\label{eq69}
\end{equation}
We would assume that there exists another operator $\hat{L}^{\prime}_{j}$ 
that also satisfies the relations (\ref{eq61}) and (\ref{eq66}) but is 
different from Eq.~(\ref{eq69}):
\begin{equation*}
\left[\hat{L}_{j}-\hat{L}_{j}^{\prime},\hat{x}_k \right]=0,
\hspace{18pt}
\left[\hat{L}_{j}-\hat{L}_{j}^{\prime},\hat{p}_k \right]=0.
\end{equation*}
From the discussion at the end of Sec.~\ref{sec4}, 
any operator that commutes with the translation generator must be a 
function only of the translation generator. 
Similarly, any operator that commutes with the position operator must be a 
function only of the position operator. Therefore the operator 
$\hat{L}_{j}-\hat{L}_{j}^{\prime}$ should be a constant. This leads to the 
conclusion that the rotation generator is given by Eq.~(\ref{eq69}), 
since a constant term adds only a physically meaningless phase to the rotation 
operator. The explicit expression for the rotation generator (\ref{eq69}) 
gives the commutation relations between the components of the rotation 
generator
\begin{equation}
\left[\hat{L}_{j},\hat{L}_{k}\right]=i\theta\varepsilon_{jkl}\hat{L}_{l},
\label{eqCRLL}
\end{equation}
which show that the rotation generator is also a vector operator. 

\section{\label{sec6}GALILEAN TRANSFORMATION}

We denote the Galilean transformation operator as $\hat{U}_{G}({\bf V},t)$. 
The position operator $\hat{{\bf r}}$ of a particle is transformed into $\hat{{\bf r}}^{\prime}$ by a Galilean transformation:
\begin{equation}
\hat{{\bf r}}
\rightarrow
\hat{{\bf r}}^{\prime}=
\hat{U}_{G}^{\dag}({\bf V},t)\hat{{\bf r}}
\hat{U}_{G}({\bf V},t)=\hat{{\bf r}}+{\bf V}t.
\label{eq72}
\end{equation}
The Galilean transformation operator must be unitary. 

It is easily seen from the property of the Galilean transformation that the 
expression
\begin{equation}
\hat{U}_{G}({\bf V},t)\hat{U}_{G}({\bf V}^{\prime},t)
=
\hat{U}_{G}({\bf V}^{\prime},t)\hat{U}_{G}({\bf V},t)
=
\hat{U}_{G}({\bf V}+{\bf V}^{\prime},t)
\label{eq73}
\end{equation}
is valid. Like the translation operator, therefore, the Galilean transformation 
operator can be expressed as
\begin{equation}
\hat{U}_{G}({\bf V},t)
=
\exp\left[-\frac{i}{\theta}{\bf V}\cdot\hat{{\bf G}}(t)\right]
\label{eq74}
\end{equation}
where $\hat{{\bf G}}(t)$ is a self-adjoint vector operator and is called the \textit{Galilean transformation generator}. 
It is also required from Eq.~(\ref{eq73}) that the relations
\begin{equation}
\left[\hat{G}_{j},\hat{G}_{k}\right]=0
\label{eq75}
\end{equation}
hold. It follows from Eqs.~(\ref{eq72}), (\ref{eq74}), and (\ref{eq75}) that
\begin{equation*}
\left[\hat{x}_{j},\hat{G}_{k}\right]=i\theta t\delta_{jk}.
\end{equation*}
We therefore obtain, with the help of the commutation relation (\ref{eqccr3}), 
\begin{equation*}
\left[\hat{x}_{j},\hat{G}_{k}-t\hat{p}_{k}\right]=0.
\end{equation*}

This equation gives that
\begin{equation}
\hat{G}_{j}
=
t \hat{p}_{j}+ \hat{f}_{j}(\hat{{\bf r}}),
\label{eq78}
\end{equation}
where $\hat{{\bf f}}(\hat{{\bf r}})$ is a vector function of 
$\hat{{\bf r}}$, since any operator that commutes with the position 
operator should be a function only of the position operator. 
The Galilean transformation for the translation generator can then 
be expanded as
\begin{eqnarray}
\hat{U}_{G}^{\dag}\hat{p}_{j}\hat{U}_{G}
&=& \sum_{n=0}^{\infty} \frac{(-i)^n}{n!}\left[ \left[ \cdots \left[ \left[ 
\hat{p}_{j},\hat{G} \right] , \hat{G} \right] , \cdots \right] , \hat{G} 
\right] \nonumber \\
&=& \hat{p}_{j}-V_{k}
\displaystyle
\frac{\partial\hat{f}_{k}}{\partial\hat{x}_{j}}-\frac{t}{2}V_{k}V_{l}
\displaystyle
\frac{\partial^{2}\hat{f}_{k}}{\partial\hat{x}_{j}\partial\hat{x}_{l}}+\cdots,
\label{eq80}
\end{eqnarray}
where we have put
\begin{equation*}
\hat{G}\equiv\frac{1}{\theta}{\bf V}\cdot\hat{{\bf G}}=
\frac{1}{\theta}\left(t{\bf V}\cdot\hat{{\bf p}}+
{\bf V}\cdot\hat{{\bf f}}\right).
\end{equation*}

The transformed translation generator should depend neither on the position 
operator nor on time, because of principle of space and time requiring that space and time are homogeneous. 
Terms of higher than first order in the velocity in the right side 
of Eq.~(\ref{eq80}) should therefore vanish, so that $\hat{{\bf f}}$ must 
a function linear in the position operator:
\begin{equation}
\hat{{\bf f}}(\hat{{\bf r}})
=
- m \hat{{\bf r}},
\label{eq81}
\end{equation}
where $m$ is a constant. The operator $\hat{{\bf f}}$ can include a 
constant term, but it can be omitted, since it only adds a physically 
meaningless phase to the Galilean transformation operator. 
Thus the translation generator is transformed by the Galilean transformation as
\begin{equation}
\hat{U}_{G}^{\dag}\hat{\bf p}\hat{U}_{G}=\hat{\bf p} + m {\bf V}.
\label{eq82}
\end{equation}
This result might be expected if the translation generator is identified with momentum operator, and then the constant $m$ can be identified with mass. 
We, however, have not identified the translation generator with momentum operator as yet, but shall do in the next section.
It will be shown in Sec.~\ref{sec8} that the constant $m$ be mass of the particle. 

Substituting (\ref{eq81}) into (\ref{eq78}), we obtain the explicit expression 
for the Galilean transformation generator:
\begin{equation}
\hat{{\bf G}}=t\hat{{\bf p}}- m \hat{{\bf r}}
\label{eqGTG}
\end{equation}
Usually the Galilean transformation generator is obtained from the transformation rules for both position and velocity. 
We cannot adopt such the way since we have not yet defined velocity operator. 
It is important that the Galilean transformation generator has been obtained without the help of the transformation rule for velocity.
Thus the Galilean transformation operator is expressed as
\begin{equation*}
\hat{U}_{G}({\bf V},t)=
\exp\left[-\frac{i}{\theta}{\bf V}\cdot\hat{{\bf G}}\right]
=\exp\left[-\frac{i}{\theta}{\bf V}\cdot
(t\hat{{\bf p}}- m \hat{{\bf r}})\right].
\end{equation*}
The operator $\hat{U}_{G}$ can also be rewritten as
\begin{eqnarray}
\hat{U}_{G}({\bf V},t)
&=&
\exp\left(-i\frac{m V^{2}}{2\theta}t\right)
\exp\left(i\frac{m{\bf V}}{\theta}\cdot\hat{{\bf r}}\right)
\exp\left(-i\frac{{\bf V}t}{\theta}\cdot\hat{{\bf p}}\right)
\nonumber
\\[6pt]
&=&
\exp\left(i\frac{mV^{2}}{2\theta}t\right)
\exp\left(-i\frac{{\bf V}t}{\theta}\cdot\hat{{\bf p}}\right)
\exp\left(i\frac{m{\bf V}}{\theta}\cdot\hat{{\bf r}}\right).
\label{eq85}
\end{eqnarray}

\section{\label{sec7}LAGRANGIAN AND EQUATION OF MOTION FOR A FREE PARTICLE}

In this section we shall obtain the explicit expression for the Lagrangian 
for a free particle on the basis of the principles introduced in Sec.~\ref{sec2}. The Lagrangian must be represented by state vectors 
in the present case, since it should be in general represented by variables 
that represent states of a system of interest and states of a microscopic particle are represented by state vectors. Thus the Lagrangian must be represented in terms of inner products in the state vector space, since it should be a real c-number. According to the wave principle, the Lagrangian must be linear in state kets and antilinear in state bras. If it would not be the case, the equation of motion derived from it would not satisfy the principle of superposition.

\subsection{\label{sec71}Requirements of the principle of space and time}

The Lagrangian must be independent of time and invariant with respect to space translation and rotation due to the principle of space and time.
First, since the Lagrangian must be independent of time, we obtain a general expression for the Lagrangian as
\begin{eqnarray}
L&=&\langle\psi\vert\hat{C}_{0}\vert\psi\rangle+\frac{i}{2}
\left[\langle\psi\vert\hat{C}_{1}
\displaystyle
\frac{d\vert\psi\rangle}{dt}-\frac{d\langle\psi\vert}{dt}\hat{C}_{1}
\vert\psi\rangle\right]
\nonumber \\[6pt]
&+&\frac{d\langle\psi\vert}{dt}\hat{C}_{2}\frac{d\vert\psi\rangle}{dt}
+\frac{i}{2}\left[
\frac{d\langle\psi\vert}{dt}\hat{C}_{3}\frac{d^{2}\vert\psi\rangle}{dt^{2}}
-\frac{d^{2}\langle\psi\vert}{dt^{2}}\hat{C}_{3}\frac{d\vert\psi\rangle}{dt}
\right]+\cdots
\label{eq87}
\end{eqnarray}
where all $\hat{C}_{n}$ are self-adjoint operators which are independent of 
time. The above equation does not involve such terms as, for example,
\begin{equation*}
\langle\psi\vert\hat{A}
\displaystyle
\frac{d\vert\psi\rangle}{dt}+\frac{d\langle\psi\vert}{dt}\hat{A}
\vert\psi\rangle=\frac{d}{dt}\langle\psi\vert\hat{A}\vert\psi\rangle,
\end{equation*}
since the terms that are exact differentials with respect to time do not contribute the equation of 
motion.\footnote{See L.D. Landau and E.M. Lifshitz, ``\textit{Mechanics}'' (Course of Theoretical Physics, Vol. 1) 3rd ed. (Pergamon Press, New York, 1976), Sec.2.}

Next, we consider the invariance with respect to space translation. Suppose that a state vector is transformed by a space translation as
\begin{equation*}
\vert\psi\rangle\rightarrow\vert\psi^{\prime}\rangle=\hat{U}({\bf a})
\vert\psi\rangle.
\end{equation*}
Since the Lagrangian is invariant with respect to this transformation, all the 
operators that are involved in the Lagrangian must be unchanged by the 
transformation:
\begin{equation*}
\hat{U}^{\dag}({\bf a})\hat{C}_{n}\hat{U}({\bf a})=\hat{C}_{n},
\end{equation*}
which follows that
\begin{equation*}
\left[\hat{C}_{n},\hat{{\bf p}}\right]=0.
\end{equation*}
This means that $\hat{C}_{n}$ are functions only of the translation generator. 
In similar fashion, it is shown from space-rotation invariance that $\hat{C}_{n}$ 
must commute with the rotation generator:
\begin{equation*}
\left[\hat{C}_{n},\hat{{\bf L}}\right]=0.
\end{equation*}
Thus $\hat{C}_{n}$ are scalars that are dependent only on the translation generator, so that $\hat{C}_{n}$ can in general be expressed as
\begin{equation}
\hat{C}_{n}=c_{n0}+c_{n2}\hat{{\bf p}}^{2}+
c_{n4}(\hat{{\bf p}}^{2})^{2}+\cdots
\label{defCn}
\end{equation}

From Hamilton's principle we obtain the Lagrangian equations of motion
\begin{eqnarray*}
 \frac{{\delta L}}{{\delta \left| \psi  \right\rangle }} - \frac{d}{{dt}}\frac{{\delta L}}{{\delta \left| {\dot \psi } \right\rangle }} + \frac{{d^2 }}{{dt^2 }}\frac{{\delta L}}{{\delta \left| {\ddot \psi } \right\rangle }} -  \cdots  = 0 \\ 
 \frac{{\delta L}}{{\delta \left\langle \psi  \right|}} - \frac{d}{{dt}}\frac{{\delta L}}{{\delta \left\langle {\dot \psi } \right|}} + \frac{{d^2 }}{{dt^2 }}\frac{{\delta L}}{{\delta \left\langle {\ddot \psi } \right|}} -  \cdots  = 0 
\end{eqnarray*}
where we have adopted the abbreviation
\begin{equation*}
\vert\dot{\psi}\rangle\equiv\frac{d}{dt}\vert\psi\rangle,
\hspace{34pt}
\vert\ddot{\psi}\rangle\equiv\frac{d^2}{dt^2}\vert\psi\rangle,
\hspace{34pt}
\mathrm{etc.}
\end{equation*}
Higher order time derivatives can appear in the equations of motion, since the Lagrangian may involve higher order time derivatives of state vectors.

From the Lagrangian (\ref{eq87}) we can obtain the equation of motion
\begin{equation*}
i\hat{C}_{1}\frac{d}{dt}\vert\psi\rangle-
\hat{C}_{2}\frac{d^{2}}{dt^{2}}\vert\psi\rangle-
i\hat{C}_{3}\frac{d^{3}}{dt^{3}}\vert\psi\rangle-
\cdots+\hat{C}_{0}\vert\psi\rangle=0.
\end{equation*}
In the interests of simplicity, we rewrite the above equation as
\begin{equation}
ic_{10}\frac{d}{dt}\vert\psi\rangle=
-\hat{C}_{0}\vert\psi\rangle-
i\hat{C}^{\prime}_{1}\frac{d\vert\psi\rangle}{dt}+
\hat{C}_{2}\frac{d^{2}\vert\psi\rangle}{dt^{2}}+
i\hat{C}_{3}\frac{d^{3}\vert\psi\rangle}{dt^{3}}\cdots
\label{eq97}
\end{equation}
where $c_{10}$ is a real constant and we have put
\begin{equation}
\hat{C}_{1}^{\prime}\equiv\hat{C}_{1}-c_{10}=c_{12}\hat{{\bf p}}^{2}+
c_{14}(\hat{{\bf p}}^{2})^{2}+\cdots.
\label{defC'1}
\end{equation}

\subsection{\label{sec72}Requirements of the probability principle \hspace{10pt} --- Derivation of Schr\"odinger equation ---}

In Sec.~\ref{sec3} it has been shown from the probability principle and the principle of indestructibility and increatibility of particles
that the wave function must satisfy the equation of continuity (\ref{eq15}). 
We therefore require 
that the equation of continuity is derived from the equation of motion for the 
state vector.
The time derivative of the position 
probability density of the particle can be rewritten as
\[
\frac{\partial}{\partial t}\vert\psi({\bf r})\vert^{2}
=
\frac{\partial}{\partial t}\langle\psi\vert{\bf r}\rangle
\langle{\bf r}\vert\psi\rangle
=
\langle\psi\vert{\bf r}\rangle\langle{\bf r}\vert
\frac{d\vert\psi\rangle}{dt}
+
\frac{d\langle\psi\vert}{dt}
\vert{\bf r}\rangle\langle{\bf r}\vert\psi\rangle.
\]
Substitution of Eq.~(\ref{eq97}) into this equation gives
\begin{eqnarray}
ic_{10}\frac{\partial}{\partial t}\vert\psi({\bf r})\vert^{2}
&=&\langle\psi\vert{\bf r}\rangle\langle{\bf r}\vert
\left[-\hat{C}_{0}\vert\psi\rangle-
i\hat{C}^{\prime}_{1}\frac{d\vert\psi\rangle}{dt}+
\hat{C}_{2}\frac{d^{2}\vert\psi\rangle}{dt^{2}}+\cdots\right]
\nonumber \\[8pt]
&-&\left[-\langle\psi\vert\hat{C}_{0}+
i\frac{d\langle\psi\vert}{dt}\hat{C}^{\prime}_{1}+
\frac{d^{2}\langle\psi\vert}{dt^{2}}\hat{C}_{2}+\cdots\right]
\vert{\bf r}\rangle\langle{\bf r}\vert\psi\rangle.
\label{eq100}
\end{eqnarray}
This reduces to the equation of continuity, if and only if
\begin{eqnarray*}
\hat{C}^{\prime}_{1}=0,
\hspace{30pt} \mathrm{and} \hspace{30pt}
\hat{C}_{n}=0 \hspace{20pt} \mathrm{for} \hspace{10pt} n\geq2
\end{eqnarray*}
(see Appendix). 
Because of the requirement that the equation of continuity derives 
from the equation of motion for the state vector, we conclude that the 
equation of motion must have the form
\begin{equation*}
ic_{10}\frac{d}{dt}\vert\psi\rangle=-\hat{C}_{0}\vert\psi\rangle
\end{equation*}

The Lagrangian, therefore, must be given by
\begin{equation}
L
=
\frac{ih}{4\pi}\left[\langle\psi\vert\frac{d\vert\psi\rangle}{dt}
-
\frac{d\langle\psi\vert}{dt}\vert\psi\rangle\right]-
\langle\psi\vert\hat{H}\vert\psi\rangle,
\label{eqLag1}
\end{equation}
where we have put
\begin{equation*}
\hat{H}
=
- \hat{C}_{0},
\hspace{28pt}
\frac{h}{2\pi}
=
c_{10}
\end{equation*}
The operator $\hat H$ is the Hamiltonian.
It shall be shown in Sec.~\ref{sec8} that the Hamiltonian is the operator corresponding to energy.
Its explicit form shall be derived below. 
The quantity $h$ is now an undetermined constant, which shall be shown to be equal to Planck's constant in the next section. 
As an alternative expression for the Lagrangian, we can take
\begin{equation}
L=\langle\psi\vert i\frac{h}{2\pi}\frac{d}{dt}-\hat{H}\vert\psi\rangle,
\label{eqLag2}
\end{equation}
which is obtained by adding a exact differential with respect to time, $\frac{ih}{4\pi} \frac{d}{dt} \langle\psi\vert\psi\rangle$, to the 
expression (\ref{eqLag1}) and is complex rather than real. This expression for 
the Lagrangian is simple and convenient, from which we can always derive the same results as those derived from Eq.~(\ref{eqLag1}), so that it will be used henceforth.

 Since the Lagrangian involves only the first order time derivatives of the state vector, the Lagrangian equations of motion reduce to
\begin{equation}
\frac{d}{dt}\frac{{\delta L}}{{\delta \left| {\dot \psi } \right\rangle }} 
=
 \frac{{\delta L}}{{\delta \left| \psi  \right\rangle }},
\hspace{35pt}
\frac{d}{{dt}}\frac{{\delta L}}{{\delta \left\langle {\dot \psi } \right|}} = \frac{{\delta L}}{{\delta \left\langle \psi  \right|}}.
\label{eqLEML}
\end{equation}
The latter gives the equation of motion for the ket in the explicit expression:
\begin{equation}
i\frac{h}{2\pi}\frac{d}{dt}\vert\psi\rangle
=
\hat{H}\vert\psi\rangle.
\label{eq105}
\end{equation}
Thus we have derived Schr\"odinger equation, but have not determined the value of the constant $h$ and the form and meaning of $\hat H$ as yet.

\subsection{\label{sec73}Galilean invariance \hspace{10pt} --- Determination of the more explicit form of Hamiltonian ---}

In this subsection we shall obtain the more explicit form of the Hamiltonian $\hat{H}$ from Galilean invariance. 

The state vector is transformed by the Galilean transformation as
\begin{equation*}
\vert\psi\rangle\rightarrow\vert\psi^{\prime}\rangle=
\hat{U}_{G}({\bf V},t)\vert\psi\rangle=\exp\left[
-\frac{i}{\theta}{\bf V}\cdot(t\hat{{\bf p}}-
m \hat{{\bf r}})\right]\vert\psi\rangle.
\end{equation*}
The principle of Galilean relativity requires that the Lagrangian be invariant 
with respect to Galilean transformation:
\begin{eqnarray*}
L^{\prime}
&=&
\langle\psi^{\prime}\vert
 \frac{ih}{2\pi}\frac{d}{dt} - \hat{H} \vert\psi^{\prime}\rangle
\nonumber \\[10pt]
&=&
\langle\psi\vert
 \frac{ih}{2\pi}\frac{d}{dt} - \hat{H} \vert\psi\rangle
=
L,
\end{eqnarray*}
so that the following equation must hold:
\begin{equation*}
\frac{{ih}}{{2\pi }}\hat U_G^\dag  \left( {{\bf{V}},t} \right)\frac{{d\hat U_G \left( {{\bf{V}},t} \right)}}{{dt}} - \hat U_G^\dag  \left( {{\bf{V}},t} \right)\hat H\hat U_G \left( {{\bf{V}},t} \right) =  - \hat H .
\end{equation*}
Since Eqs.~(\ref{eq85}) and (\ref{eq82}) give the equations
\begin{equation*}
\frac{{ih}}{{2\pi }}\hat U_G^\dag  \left( {{\bf{V}},t} \right)\frac{{d\hat U_G \left( {{\bf{V}},t} \right)}}{{dt}} 
=
 \frac{h}{{2\pi \theta }}\left( \frac{1}{2}mV^2  + {\bf{V}} \cdot {\bf{\hat p}} \right)
\end{equation*}
\begin{equation*}
\hat{U}_{G}^{\dag}({\bf V},t)\hat{H}(\hat{{\bf p}})
\hat{U}_{G}({\bf V},t)=\hat{H}(\hat{{\bf p}}+m{\bf V}),
\end{equation*}
the Lagrangian is invariant with respect to Galilean transformation if
\begin{equation*}
\hat{H}(\hat{{\bf p}}+ m {\bf V})=\hat{H}(\hat{{\bf p}})
+
\frac{h}{2\pi\theta}\left(\frac{1}{2}mV^2 +
{\bf V}\cdot\hat{{\bf p}}\right).
\end{equation*}
This means that the Hamiltonian must have the form
\begin{equation}
\hat{H}
=
\frac{h}{4\pi\theta m}\hat{{\bf p}}^{2}.
\label{eqHam1}
\end{equation}
The Hamiltonian can include a real constant term, but it is neglected since it is not physically meaningful.

The Lagrangian and the Hamiltonian include three undetermined constants, $h$, $\theta$, and $m$. They will be determined bellow.

\section{\label{sec8}SYMMETRIES AND CONSERVATIVE QUANTITIES}

In this section, on the basis of conservation laws of energy, momentum, and angular momentum that are derived from principle of space and time, we shall show that the Hamiltonian is the operator corresponding to energy and that the translation generator can be identified with momentum operator.  We shall also prove that $h$ is Planck's constant and $m$ is the mass of a particle.

\subsection{\label{sec81}Homogeneity of time and conservation law of energy}

Now we discuss the conservation law of energy that is derived from 
homogeneity of time, which is required from principle of space and time. The time derivative of the Lagrangian can be expressed as
\begin{equation*}
\frac{dL}{dt}=\frac{\partial L}{\partial t}+\frac{d}{dt}
\left(
\frac{\delta L}{\delta\vert\dot{\psi}\rangle}\vert\dot{\psi}\rangle+
\langle\dot{\psi}\vert\frac{\delta L}{\delta\langle\dot{\psi}\vert}
\right),
\end{equation*}
with the help of the Lagrangian equations of motion (\ref{eqLEML}).
It follows from the above equation that
\begin{equation*}
\frac{\partial L}{\partial t}
=
-\frac{d}{dt}
\left(
\frac{\delta L}{\delta\vert\dot{\psi}\rangle}\vert\dot{\psi}\rangle+
\langle\dot{\psi}\vert\frac{\delta L}{\delta\langle\dot{\psi}\vert}
-L
\right)
=0
\end{equation*}
since the Lagrangian does not depend on time explicitly because of 
homogeneity of time. This equation means that the quantity in the parentheses 
on the right side is a constant of the motion. 
This conserved quantity, which is energy since it is derived from 
homogeneity of time, is called the Hamiltonian in classical mechanics and is denoted by $H$. 
Then the classical Hamiltonian is expressed explicitly from Eq.~(\ref{eqLag2}) as
\begin{equation*}
H
=
\frac{\delta L}{\delta\vert\dot{\psi}\rangle}\vert\dot{\psi}\rangle+
\langle\dot{\psi}\vert\frac{\delta L}{\delta\langle\dot{\psi}\vert}
-L
=
\langle \psi \vert \hat{H} \vert \psi \rangle.
\end{equation*}
This shows that the Hamiltonian $\hat{H}$ is the operator corresponding to energy, as it should be. 

If a state vector $\vert\psi(t)\rangle$ is an eigenket of $\hat{H}$ with an 
eigenvalue $E$, then from Eq.~(\ref{eq105}) we obtain 
\begin{equation*}
i\frac{h}{2\pi}\frac{d}{dt}\vert\psi(t)\rangle=E\vert\psi(t)\rangle,
\end{equation*}
the solution of which is
\begin{equation*}
\vert\psi(t)\rangle=e^{-i2\pi Et/h}\vert\psi(0)\rangle.
\end{equation*}
The state vector is a periodic function of time and the relation between its 
frequency $\nu$ and the energy $E$ is given by
\begin{equation*}
E=h\nu,
\end{equation*}
which is one of Einstein-de Broglie formulas. This shows that the constant $h$ 
should be Planck's constant.

\subsection{\label{sec82}Homogeneity and isotropy of space and conservation 
laws of momentum and angular momentum}

Now we discuss the conservation law of momentum that is derived from 
homogeneity of space, which is required from principle of space and time. 
The conservation law of momentum that is valid for any 
branch of physics is one of fundamental laws in physics.  
Momentum must therefore be defined on the common basis of physics. 
We thus define momentum as the conserved additive vector that is derived from 
homogeneity of space or translational symmetry. 

We suppose that a translation by {\bf a} transforms 
$\vert\psi\rangle$ into $\vert\psi^{\prime}\rangle$:
\[
\left| \psi  \right\rangle  \to \left| {\psi '} \right\rangle  = \hat U\left( {\bf{a}} \right)\left| \psi  \right\rangle  = e^{ - i{{{\bf{a}} \cdot {\bf{\hat p}}} \mathord{\left/
 {\vphantom {{{\bf{a}} \cdot {\bf{\hat p}}} \theta }} \right.
 \kern-\nulldelimiterspace} \theta }} \left| \psi  \right\rangle ,
\]
where $ \hat U\left( {\bf{a}} \right) $ is given by Eq.~(\ref{eq28}).
Since the Lagrangian $L^{\prime}$, which is the Lagrangian given by 
$\vert\psi^{\prime}\rangle$, is independent of {\bf a} because of 
homogeneity of space:
\begin{equation}
\left.\frac{\partial L^{\prime}}{\partial{\bf a}}\right\vert_{{\bf a}=0}=0 .
\label{eq121}
\end{equation}
On the other hand we obtain
\begin{eqnarray}
\left.\frac{\partial L^{\prime}}{\partial{\bf a}}\right\vert_{{\bf a}=0}
&=&
\frac{\delta L}{\delta\vert\psi\rangle}
\left.\frac{\partial \vert\psi^{\prime}\rangle}{\partial{\bf a}}\right\vert_{{\bf a}=0}
+
\frac{\delta L}{\delta\vert\dot{\psi}\rangle}
\left.\frac{\partial \vert\dot{\psi}^{\prime}\rangle}{\partial{\bf a}}
\right\vert_{{\bf a}=0}
\nonumber \\[10pt]
&&+
\left.\frac{\partial \langle\psi^{\prime}\vert}{\partial{\bf a}}\right\vert_{{\bf a}=0}\frac{\delta L}{\delta\langle\psi\vert}
+
\left.\frac{\partial \langle\dot{\psi}^{\prime}\vert}{\partial{\bf a}}
\right\vert_{{\bf a}=0}\frac{\delta L}{\delta\langle\dot{\psi}\vert}
\nonumber \\[10pt]
&=&
\frac{d}{{dt}}
\left[ 
{\frac{\delta L}{\delta \vert \dot{\psi} \rangle }\left. {\frac{{\partial \left| {\psi '} \right\rangle }}{{\partial {\bf{a}}}}} \right|_{{\bf{a}} = 0}  
+
 \frac{\delta L}{\delta \langle \dot{\psi} \vert}\left. {\frac{\partial \left\langle {\psi '} \right|}{{\partial {\bf{a}}}}} \right|_{{\bf{a}} = 0} } 
\right]
\nonumber \\[10pt]
&=&
\frac{h}{2\pi\theta}\frac{d}{dt}\langle\psi\vert
\hat{{\bf p}}\vert\psi\rangle,
\label{eqConP}
\end{eqnarray}
where use has been made of Eqs.~(\ref{eqLEML}) and (\ref{eqLag2}).
It follows from Eqs.~(\ref{eq121}) and (\ref{eqConP}) that the translation 
generator is conserved under translational symmetry. Therefore the translation generator should be proportional to momentum operator.

Next, we consider a system consisting of many particles which need not be identical. It is clear that in this system the translation generator for each particle can be defined as
\begin{equation*}
\left[\hat{x}_{\alpha j},\hat{p}_{\beta k}\right]=
i\theta\delta_{\alpha\beta}\delta_{jk},\hspace{32pt}
\left[\hat{p}_{\alpha j},\hat{p}_{\beta k}\right]=0,
\end{equation*}
where Greek subscripts denote particles and alphabetic subscripts denote the 
components of vectors. The Hamiltonian for this system should be given as a function of translation generators and position operators of those particles: $ \hat H\left( {\left\{ {{\bf{\hat p}}_\alpha  } \right\};\left\{ {{\bf{\hat r}}_\alpha  } \right\}} \right) $.  The translation operator that translates the system by {\bf a} is given by
\[
\hat U\left( {\bf{a}} \right) = \exp \left[ { - i\frac{{\bf{a}}}{\theta } \cdot \sum\limits_\alpha  {{\bf{\hat p}}_\alpha  } } \right].
\]
This shows that translation generators are additive, which is also required for momenta, if and only if the constant $\theta$ involved in the translation operator is a universal constant. 
We require from the principle of space and time that the system has translational symmetry;
\[
\hat U^\dag  \left( {\bf{a}} \right)\hat H\left( {\left\{ {{\bf{\hat p}}_\alpha  } \right\};\left\{ {{\bf{\hat r}}_\alpha  } \right\}} \right)\hat U\left( {\bf{a}} \right) = \hat H\left( {\left\{ {{\bf{\hat p}}_\alpha  } \right\};\left\{ {{\bf{\hat r}}_\alpha   + {\bf{a}}} \right\}} \right) = \hat H\left( {\left\{ {{\bf{\hat p}}_\alpha  } \right\};\left\{ {{\bf{\hat r}}_\alpha  } \right\}} \right).
\]
Then we can obtain the equation
\[
\frac{d}{{dt}}\left\langle \psi  \right|\sum\limits_\alpha  {{\bf{\hat p}}_\alpha  } \left| \psi  \right\rangle  = 0,
\]
like Eq.~(\ref{eqConP}) for a free particle.  Therefore even if there exist interactions between particles, the sum of translation generators is conserved. Since the conserved quantity that is derived from translational symmetry is (total) momentum, the sum of all translation generators must be proportional to the total momentum operator. 
We therefore identify translation generators with momentum operators from now on, the condition for which will be shown to be $\theta = h / {2 \pi}$ in Sec.~\ref{Int}.

In a similar way, it is shown that the rotation generator is proportional to 
the angular momentum operator that is the conserved quantity derived from the 
isotropy of space, which is also required from the principle of space and time. 
It is apparent that the rotation generator is identical with the angular momentum operator if the translation generator is identical with the momentum operator.

\subsection{\label{secCQ}Quantum-mechanical definition of a conserved quantity}

The above discussion has shown that the conservation law in quantum mechanics is expressed such that the expectation value of the operator corresponding to a conserved dynamical variable is independent of time for any time-dependent state vector which is a solution of Schr\"odinger equation;
if
\begin{equation}
\frac{d}{dt}\langle\hat{A}\rangle=
\frac{d}{dt}\langle\psi\vert\hat{A}\vert\psi\rangle=0,
\label{eq8}
\end{equation}
where $\vert\psi\rangle$ is any time-dependent state vector that describes the considered system, then the dynamical variable $A$ is conserved. 
This is important because it should be characteristic of a physical system and independent of its state that a certain dynamical variable of the system is conserved. 

With the help of Schr\"odinger equation (\ref{eq105}), the time derivative of the expectation value of any operator $\hat{A}$ is given by
\begin{equation}
\frac{d}{dt}\langle\hat{A}\rangle
= \frac{d}{dt}\langle\psi\vert\hat{A}\vert\psi\rangle
= \langle\psi\vert\frac{\partial\hat{A}}{\partial t}\vert\psi\rangle
+ \frac{2\pi}{ih}\langle\psi\vert\left[\hat{A},\hat{H}\right]
\vert\psi\rangle.
\label{eq125}
\end{equation}
Thus the definition of a conserved quantity (\ref{eq8}) can be rewritten as
\begin{equation*}
\left[\hat{A},\hat{H}\right]=0,
\end{equation*}
if $\hat{A}$ does not depend explicitly on time. This is an alternative expression of the definition of the conserved quantity in quantum mechanics. Invariance of the Hamiltonian 
with respect to translation and rotation gives the relations
\begin{equation*}
\left[\hat{{\bf p}},\hat{H}\right]=0,
\end{equation*}
\begin{equation*}
\left[\hat{{\bf L}},\hat{H}\right]=0,
\end{equation*}
which indicate that the momentum and angular momentum are constants of the motion.

\subsection{\label{sec84}Velocity operator and mass}

Since we deal with an object that is classically considered to be a particle, 
we naturally introduce mass as its attribute. 
In classical mechanics, mass is usually introduced as a 
proportionality constant between force and acceleration. However, we cannot 
adopt this definition of mass, since the concepts of acceleration and force do 
not have definite meanings in quantum mechanics. 
It is reasonable to define the mass of a particle as the proportionality constant between its momentum and velocity operators
\footnote{Both of mass and force are defined by equation of motion in classical mechanics, so there is an ambiguity. Instead, it is valid to define mass as a proportionality between the momentum that is a constant of motion derived from homogeneity of space and the velocity that is obviously defined as the time derivative of the position.},
since the momentum operator is proportional to the velocity 
operator in quantum mechanics as well as in classical mechanics, as shall be 
shown sooner.

It is reasonable that we define the velocity operator as the operator that has 
the expectation value equal to the time derivative of the expectation value of 
the position operator:
\begin{equation*}
\langle\hat{{\bf v}}\rangle\equiv\frac{d}{dt}
\langle\hat{{\bf r}}\rangle.
\end{equation*}
The definition of the velocity operator can be rewritten as
\begin{equation*}
\hat{{\bf v}}\equiv\frac{2\pi}{ih}
\left[\hat{{\bf r}},\hat{H}\right],
\end{equation*}
since the time derivative of the expectation value is generally given by 
Eq.~(\ref{eq125}). Substitution of Eq.~(\ref{eqHam1}) into the definition of the velocity operator gives
\[
\hat{{\bf p}}= m \hat{{\bf v}}
\]
with the help of the commutation relations (\ref{eqccr3}). 
We have shown that the velocity and momentum operators, which are defined 
independently of each other, are proportional to each other. 
From our definition of mass, it has verified that the constant $m$, which was introduced in the discussion of Galilean transformation of Sec.~\ref{sec6}, is the mass of the particle. 
From Eq.~(\ref{eq82}), therefore, we obtain the Galilean transformation for the velocity operator:
\[
{\bf{\hat v}} \to {\bf{\hat v'}} = \hat U_G ^\dag  {\bf{\hat v}}\hat U_G  = {\bf{\hat v}} + {\bf{V}},
\]
as should be.

\section{\label{Int}A SYSTEM CONSISTING OF TWO INTERACTING PARTICLES AND A PARTICLE IN AN EXTERNAL FIELD}

If there exists an external field, Galilean principle of relativity and homogeneity of space are broken, and further homogeneity of time is also broken if the external field depends on time. 
This fact means that it is difficult to obtain the properties of the external field on the basis of the principles.
The motion in an external field, however, can be regarded as the relative motion of two particles that interact with each other. 
Thus in this section we consider a system that consists of two interacting particles; the six principles should naturally hold for the system, including Galilean principle of relativity and principle of space and time. 
We investigate to what extent the properties of the interaction between two particles are restricted from the principles. 
Then we obtain the Lagrangian for a particle in an external field as that for the relative motion in the two-particle system. 
Furthermore, by applying the conservation law of momentum to the system that consists of a quantum particle and a classical particle, we determine the last undetermined constant $\theta $. 
Thus we shall obtain the final forms of the Lagrangian and the Hamiltonian for a particle in an external field, and then we shall finally attain to the true Schr\"odinger equation.

\subsection{\label{Free2}A system consisting of two free particles}

The Lagrangian for two free particles, $L_{1+2}$, is given from additivity of Lagrangians by
\[
L_{1 + 2}  = L_1  + L_2  = \left\langle {\psi _1 } \right|\frac{{ih_1 }}{{2\pi }}\frac{d}{{dt}} - \hat H_1 \left| {\psi _1 } \right\rangle  + \left\langle {\psi _2 } \right|\frac{{ih_2 }}{{2\pi }}\frac{d}{{dt}} - \hat H_2 \left| {\psi _2 } \right\rangle ,
\]
where the Hamiltonian for the $k$-th particle $\hat H_k$ is expressed as
\[
\hat H_k  = \frac{{h_k }}{{4\pi \theta m_k }}{\bf{\hat p}}_k ^2 ,
\]
and $ m_k $ and $h_k$ are the mass and Planck' constant of the $k$-th particle, respectively. We have assumed that Planck's constant might not be a universal constant. 
A state vector $ \left| {\psi _{1 + 2} } \right\rangle $ for the system consisting of two free particles should be given by a direct product of each state vector for two particles, $ \left| {\psi _1 } \right\rangle  \otimes \left| {\psi _2 } \right\rangle $. 
The Lagrangian $L_{1+2}$ should therefore be expressed in terms of $ \left| {\psi _1 } \right\rangle  \otimes \left| {\psi _2 } \right\rangle $:
\begin{equation}
\begin{array}{l}
 L_{1 + 2}  = \left\langle {\psi _1 } \right| \otimes \left\langle {\psi _2 } \right| \cdot \frac{{ih_1 }}{{2\pi }}\frac{{d\left| {\psi _1 } \right\rangle }}{{dt}} \otimes \left| {\psi _2 } \right\rangle  + \left\langle {\psi _1 } \right| \otimes \left\langle {\psi _2 } \right| \cdot \left| {\psi _1 } \right\rangle  \otimes \frac{{ih_2 }}{{2\pi }}\frac{{d\left| {\psi _2 } \right\rangle }}{{dt}} \\ 
 \quad \quad \quad \quad \quad \quad  - \left\langle {\psi _1 } \right| \otimes \left\langle {\psi _2 } \right|\hat H_1  + \hat H_2 \left| {\psi _1 } \right\rangle  \otimes \left| {\psi _2 } \right\rangle
 \end{array}
\label{L2F}
\end{equation}

For a system that consists of two interacting particles, however, it is evident that its state vector cannot be expressed as a direct product. 
Thus the Lagrangian for two interacting particles should be expressed using a state vector $ \left| {\psi _{1 + 2} } \right\rangle $, which requires
\[
h_1  = h_2 .
\]
If it would not be the case, the time derivatives in Eq.~(\ref{L2F}) could not be expressed by $ \left| {\psi _{1 + 2} } \right\rangle $ and $ \left\langle {\psi _{1 + 2} } \right| $.
It has been verified that Planck's constant should be a universal constant. Thus the Lagrangian for two noninteracting particles is given by
\[
L_{1 + 2}  = \left\langle {\psi _{1 + 2} } \right|\frac{{ih}}{{2\pi }}\frac{d}{{dt}} - \hat H_1  - \hat H_2 \left| {\psi _{1 + 2} } \right\rangle,
\]
where the Hamiltonian $ \hat{H}_k $ is expressed as
\[
\hat H_k  = \frac{h}{{4\pi \theta m_k }}{\bf{\hat p}}_k ^2 .
\]

\subsection{\label{EC2}Property of a two-particle wave function}

In order to investigate the physically permitted form of the Lagrangian for two interacting particles we consider the property of a two-particle wave function that is required to satisfy the equation of continuity for the position probability density of each particle which is derived from the two-particle wave function.

The two-particle wave function $ \psi _{1 + 2} \left( {{\bf{r}}_1 ,{\bf{r}}_2 } \right) $ is given by
\[
\psi _{1 + 2} \left( {{\bf{r}}_1 ,{\bf{r}}_2 } \right) = \left\langle {{{\bf{r}}_1 ,{\bf{r}}_2 }}
 \mathrel{\left | {\vphantom {{{\bf{r}}_1 ,{\bf{r}}_2 } {\psi _{1 + 2} }}}
 \right. \kern-\nulldelimiterspace}
 {{\psi _{1 + 2} }} \right\rangle .
\]
The quantity
\[
\left| {\psi _{1 + 2} \left( {{\bf{r}}_1 ,{\bf{r}}_2 } \right)} \right|^2 dV_1 dV_2 
\]
gives the probability of finding particle 1 in a volume element $dV_1$ about ${\bf{r}}_1 $ and particle 2 in a volume element $dV_2$ about ${\bf{r}}_2$. Integration of it over ${\bf{r}}_2$ (${\bf{r}}_1$) gives the probability of finding particle 1 (2). Thus if we integrate the equation for the time derivative of the square of the magnitude of the two-particle wave function
\begin{equation}
\frac{\partial }{{\partial t}}\left| {\psi _{1 + 2} \left( {{\bf{r}}_1 ,{\bf{r}}_2 } \right)} \right|^2  = F\left( {\{ \psi _{1 + 2} \left( {{\bf{r}}_1 ,{\bf{r}}_2 } \right)\} } \right)
\label{EC21}
\end{equation}
over ${\bf{r}}_2$ (${\bf{r}}_1$), we should obtain the equation of continuity for particle 1 (2). This requires that the equation (\ref{EC21}) is of the form
\begin{equation}
\frac{\partial }{{\partial t}}\left| {\psi _{1 + 2} \left( {{\bf{r}}_{1,} {\bf{r}}_2 } \right)} \right|^2  =  - \nabla _1  \cdot {\bf{J}}_1  - \nabla _2  \cdot {\bf{J}}_2  - \frac{{\partial ^2 }}{{\partial x_{1k} \partial x_{2l} }}T_{kl} .
\label{EC22}
\end{equation}
We call this equation the equation of continuity for two particles expedientially.

\subsection{\label{Int2}Interacting two particles}

The Lagrangian for a system that consists of two interacting particles can be expressed in general as
\[
L_{1 + 2}  = \left\langle {\psi _{1 + 2} } \right|\frac{{ih}}{{2\pi }}\frac{d}{{dt}} - \hat H_1  - \hat H_2  - \hat V\left| {\psi _{1 + 2} } \right\rangle .
\]
We assume that the interaction $\hat V$ has the form
\[
\hat V = \hat V\left( {\frac{d}{{dt}},{\bf{\hat r}}_1  - {\bf{\hat r}}_2 ,{\bf{\hat p}}_1 ,{\bf{\hat p}}_2 } \right)
\]
so that it satisfies homogeneity of space and time.

We now investigate whether the interaction is permitted to involve time differential operators. 
If the interaction involves second-order or higher time differential operators, or the first-order time differential operator with a factor that depends on ${\bf{\hat r}}_1  - {\bf{\hat r}}_2 $, ${\bf{\hat p}}_1 $ or $\hat{\bf p}_2$, then Eq.~(\ref{EC22}) does not hold. 
This is easily verified by replacing $\left| \psi  \right\rangle $ with $\left| {\psi _{1 + 2} } \right\rangle $ in the discussion in Appendix. If the interaction involves first-order time derivative with a constant factor, it effects changing of Planck's constant. This is not permitted. 
If it would be the case, a three-particle system that consists of two-interacting particles and a free particle would not satisfy the additivity of Lagrangians. 
Thus the interaction must not involve time differential operators.

We next consider invariance under Galilean transformation. The Galilean transformation operator for a two-particle system is given by
\[
\hat U_G \left( {{\bf{v}},t} \right) = \exp \left[ { - \frac{i}{\theta }{\bf{v}} \cdot \left\{ {\left( {t{\bf{\hat p}}_1  - m_1 {\bf{\hat r}}_1 } \right) + \left( {t{\bf{\hat p}}_2  - m_2 {\bf{\hat r}}_2 } \right)} \right\}} \right] .
\]
The requirement that the Lagrangian be invariant under the Galilean transformation gives the equation
\[
\hat U_G ^\dag  \left( {{\bf{v}},t} \right)\hat V\hat U_G \left( {{\bf{v}},t} \right) = \hat V,
\]
from which the interaction becomes of the form
\[
\hat V = \hat V\left( {{\bf{\hat r}}_1  - {\bf{\hat r}}_2 ,\frac{{{\bf{\hat p}}_1 }}{{m_1 }} - \frac{{{\bf{\hat p}}_2 }}{{m_2 }}} \right) .
\]
Then we can write the interaction as in the following:
\begin{eqnarray*}
\hat V &=& \hat V_0 \left( {{\bf{\hat r}}_1  - {\bf{\hat r}}_2 } \right)
\nonumber \\[8pt]
&\quad& {} +
a_{20} \left( {\frac{{{\bf{\hat p}}_1 }}{{m_1 }} - \frac{{{\bf{\hat p}}_2 }}{{m_2 }}} \right)^2  + c_{40} \left[ {\left( {\frac{{{\bf{\hat p}}_1 }}{{m_1 }} - \frac{{{\bf{\hat p}}_2 }}{{m_2 }}} \right)^2 } \right]^2  +  \cdots
\nonumber \\[8pt]
&\quad& {} +
a_{11} \left\{ {\left( {{\bf{\hat r}}_1  - {\bf{\hat r}}_2 } \right) \cdot \left( {\frac{{{\bf{\hat p}}_1 }}{{m_1 }} - \frac{{{\bf{\hat p}}_2 }}{{m_2 }}} \right) + \left( {\frac{{{\bf{\hat p}}_1 }}{{m_1 }} - \frac{{{\bf{\hat p}}_2 }}{{m_2 }}} \right) \cdot \left( {{\bf{\hat r}}_1  - {\bf{\hat r}}_2 } \right)} \right\}
\nonumber \\[8pt]
&\quad& {} +
a_{22} \left\{ {\left( {{\bf{\hat r}}_1  - {\bf{\hat r}}_2 } \right) \cdot \left( {\frac{{{\bf{\hat p}}_1 }}{{m_1 }} - \frac{{{\bf{\hat p}}_2 }}{{m_2 }}} \right) + \left( {\frac{{{\bf{\hat p}}_1 }}{{m_1 }} - \frac{{{\bf{\hat p}}_2 }}{{m_2 }}} \right) \cdot \left( {{\bf{\hat r}}_1  - {\bf{\hat r}}_2 } \right)} \right\}^2  +  \cdots ,
\end{eqnarray*}
where ${\hat V}_0$ is a self-adjoint scalar operator that is a function only of ${\bf{\hat r}}_1  - {\bf{\hat r}}_2 $.
Explicit calculation shows that the terms in the second and third lines in the above equation are compatible with the equation of continuity for two particles. Dependence of interaction on momenta is generally permitted. Thus it has been verified that the Hamiltonian for a two-particle system is generally given by
\begin{equation}
\hat H_{1 + 2}  
=
 \hat H_1  + \hat H_2  + \hat V 
=
 \frac{h}{{4\pi \theta m_1 }}{\bf{\hat p}}_1 ^2  + \frac{h}{{4\pi \theta m_2 }}{\bf{\hat p}}_2 ^2  + \hat V\left( {{\bf{\hat r}}_1  - {\bf{\hat r}}_2 ,\frac{{{\bf{\hat p}}_1 }}{{m_1 }} - \frac{{{\bf{\hat p}}_2 }}{{m_2 }}} \right).
\label{Hfor2}
\end{equation}

\subsection{\label{ExtF}Lagrangian for an external field}

We now derive the Lagrangian for a particle in an external field as the Lagrangian for the relative motion of two interacting particles. 
For that purpose we divide six internal degrees of freedom into two parts, those for the relative motion and those for the center-of-mass motion.

The degrees of freedom with respect to the relative motion are specified by the relative position operator
\[
\hat{\bf r} = \hat{\bf r}_1 - \hat{\bf r}_1 
\]
and the momentum operator $\hat{\bf p}$ canonically conjugate to $\hat{\bf r}$. The momentum operator $\hat{\bf p}$ should satisfy the equation
\begin{equation}
\exp \left[ {\frac{i}{\theta }{\bf{a}} \cdot {\bf{\hat p}}} \right]{\bf{\hat r}}\exp \left[ { - \frac{i}{\theta }{\bf{a}} \cdot {\bf{\hat p}}} \right] = {\bf{\hat r}} + {\bf{a}}.
\label{TG2R}
\end{equation}
We require also that $\hat{\bf p}$ is invariant under Galilean transformation. From this requirement and with the help of Eq.~(\ref{TG2R}) we obtain
\[
{\bf{\hat p}} = \frac{{m_2 }}{{m_1  + m_2 }}{\bf{\hat p}}_1  - \frac{{m_1 }}{{m_1  + m_2 }}{\bf{\hat p}}_2 .
\]

The degrees of freedom with respect to the center-of-mass motion are specified by the sum of momenta for two particles ${\bf{\hat P}} = {\bf{\hat p}}_1  + {\bf{\hat p}}_2$ and the position operator $\hat{\bf R}$ canonically conjugate to $\hat{\bf P}$. 
The position operator $\hat{\bf R}$ should satisfy the equations
\[
\exp \left[ {\frac{i}{\theta }{\bf{a}} \cdot {\bf{\hat P}}} \right]{\bf{\hat R}}\exp \left[ { - \frac{i}{\theta }{\bf{a}} \cdot {\bf{\hat P}}} \right] = {\bf{\hat R}} + {\bf{a}}
\]
and
\[
\left[ {{\bf{\hat R}},{\bf{\hat p}}} \right] = 0.
\]
The latter is necessary for the center-of-mass motion to be independent of the relative motion.
These lead to
\[
{\bf{\hat R}} = \frac{{m_1 }}{{m_1  + m_2 }}{\bf{\hat r}}_1  + \frac{{m_2 }}{{m_1  + m_2 }}{\bf{\hat r}}_2 .
\]
These operators $\hat{\bf r}$, $\hat{\bf p}$, $\hat{\bf R}$, and $\hat{\bf P}$ satisfy the following commutation relations:
\[
\begin{array}{l}
 \left[ {\hat X_j ,\hat P_k } \right] = i\theta \delta _{jk} ,\quad \quad \left[ {\hat x_j ,\hat p_k } \right] = i\theta \delta _{jk}  \\[8pt]
 \left[ {\hat X_j ,\hat p_k } \right] = 0,\quad \quad \quad \quad \left[ {\hat x_j ,\hat P_k } \right] = 0
 \end{array}
\]
The original operators are expressed in terms of the new operators as
\begin{equation}
{\bf{\hat r}}_1  = {\bf{\hat R}} + \frac{{m_2 }}{{m_1  + m_2 }}{\bf{\hat r}},\quad \quad \quad {\bf{\hat r}}_2  = {\bf{\hat R}} - \frac{{m_1 }}{{m_1  + m_2 }}{\bf{\hat r}}
\label{O2N1}
\end{equation}
and
\begin{equation}
{\bf{\hat p}}_1  = \frac{{m_1 }}{{m_1  + m_2 }}{\bf{\hat P}} + {\bf{\hat p}},\quad \quad \quad {\bf{\hat p}}_2  = \frac{{m_2 }}{{m_1  + m_2 }}{\bf{\hat P}} - {\bf{\hat p}}.
\label{O2N2}
\end{equation}

Substitution of Eqs.~(\ref{O2N1}) and (\ref{O2N2}) into Eq.~(\ref{Hfor2}) gives
\[
\hat H_{1 + 2}  
=
 \hat H_c  + \hat H_r 
=
 \frac{h}{{4\pi \theta M}}{\bf{\hat P}}^2  
+
 \frac{h}{{4\pi \theta m}}{\bf{\hat p}}^2  
+
 \hat V\left( {{\bf{\hat r}}},\bf{{\hat p}} \right),
\]
where we have put
\[
M = m_1  + m_2 ,\quad \quad \quad m = \frac{{m_1 m_2 }}{{m_1  + m_2 }},
\]
and ${\hat H}_c$ and ${\hat H}_r$ are the Hamiltonians for the center-of-mass motion and the relative motion, respectively.
It is important that the interaction depends only on the relative motion.
We can express a two-particle state vector $\left| {\psi _{1 + 2} } \right\rangle $ as a direct product of a state vector for the center-of-mass motion $\left| \Psi  \right\rangle $ and that for the relative motion $\left| \psi  \right\rangle $ :
\[
\left| {\psi _{1 + 2} } \right\rangle  = \left| \Psi  \right\rangle  \otimes \left| \psi  \right\rangle .
\]
Then the Lagrangian can be rewritten as
\begin{eqnarray*}
L_{1 + 2}  
&=&
 \left\langle \Psi  \right| \otimes \left\langle \psi  \right|\left[ {\frac{{ih}}{{2\pi }}\frac{d}{{dt}} - \hat H_{1 + 2} } \right]\left| \Psi  \right\rangle  \otimes \left| \psi  \right\rangle
\\[8pt]
&=&
 \left\langle \Psi \right|\frac{{ih}}{{2\pi }}\frac{d}{{dt}} - \hat H_c \left| \Psi \right\rangle  
+
 \left\langle \psi  \right|\frac{{ih}}{{2\pi }}\frac{d}{{dt}} - \hat H_r\left| \psi  \right\rangle
\\[8pt]
&=&
 L_c  + L_r .
\end{eqnarray*}
The two-particle Lagrangian has been divided into the Lagrangians for the relative motion and the center-of-mass motion. 
When we regard the Lagrangian for the relative motion as the Lagrangian for a particle in an external field, the latter is given by
\[
L = \left\langle \psi  \right|\frac{{ih}}{{2\pi }}\frac{d}{{dt}} - \hat H\left| \psi  \right\rangle ,
\]
where $\hat H$ is the Hamiltonian for a particle in an external field
\begin{equation}
\hat H = \frac{h}{{4\pi \theta m}}{\bf{\hat p}}^2  + \hat V\left( {{\bf{\hat r}}},\bf{\hat p} \right).
\label{eqHam2}
\end{equation}
Thus an external field can generally depend on the momentum. This is reasonable, since the Hamiltonian for a charged particle in an electromagnetic field has such a term.
The Hamiltonian for a free particle is proportional to ${\hat {\bf p}}^2$ because the Lagrangian should be invariant under Galilean transformation. However, it is not required for a particle in an external field that its Lagrangian is invariant under Galilean transformation, so that dependence of the interaction on momentum is not restricted strictly.

\subsection{\label{sec83}Condition that translation generator is identical with momentum}

In Sec.~\ref{sec8} we have verified that the translation generator is proportional to the momentum operator and have identified the former with the latter.
Now we determine the constant $\theta$ which is included in the transformation operators so that the translation generator is 
identical with the momentum operator, which also means from Eq.~(\ref{eq69}) that 
the rotation generator is identical with the angular momentum operator. 
Since momentum is defined quantitatively in classical mechanics, we should define the momentum operator in quantum mechanics consistently with the momentum in classical mechanics.  Therefore we shall introduce a system that consists of two interacting particles, one of which is a classical particle and the other a quantum particle. In the system we determine the constant $\theta$ so that the sum of both momenta is conserved.

Suppose that a classical particle of mass $M$ interacts with a 
quantum particle of mass $m$. The interaction between these particles is naturally assumed 
to be given by
\begin{equation*}
V({\bf R}-\hat{{\bf r}}),
\end{equation*}
where {\bf R} is the coordinate of the classical particle and 
$\hat{{\bf r}}$ is the position operator of the quantum particle \footnote{The interaction can be arbitrarily chosen so long as it does not 
break translational symmetry and is physically reasonable, since it is 
introduced only to determine the constant $\theta$.}. 
Then the Lagrangian for the two-particle system is given by
\begin{equation*}
L=\frac12M\dot{{\bf R}}^{2}+\langle\psi\vert\left[
i\frac{h}{2\pi}\frac{d}{dt}-\frac{h}{4\pi\theta m}\hat{{\bf p}}^{2}-
V({\bf R}-\hat{{\bf r}})\right]\vert\psi\rangle,
\end{equation*}
from which the Lagrangian equations of motion are derived:
\begin{equation}
i\frac{h}{2\pi}\frac{d}{dt}\vert\psi\rangle=
\left[\frac{h}{4\pi\theta m}\hat{{\bf p}}^{2}+
V({\bf R}-\hat{{\bf r}})\right]\vert\psi\rangle
\label{eq134}
\end{equation}
\begin{equation}
M\ddot{{\bf R}}=-\nabla_{{\mbox R}}\langle\psi\vert
V({\bf R}-\hat{{\bf r}})\vert\psi\rangle.
\label{eq135}
\end{equation}
The analytical-mechanical Hamiltonian $H$ is obtained as
\begin{eqnarray*}
H 
&=& 
{\bf P} \cdot \dot{{\bf R}} 
+ 
\langle \Pi \vert \frac{d}{dt} \vert \psi \rangle
- L
\nonumber \\[8pt]
&=& 
\frac{{\bf P}^{2}}{2M}
+
\frac{2 \pi}{ih}\langle \Pi \vert
\left[ \frac{h}{4\pi\theta m} \hat{{\bf p}}^{2}
+
V({\bf R}-\hat{{\bf r}}) \right] \vert\psi\rangle,
\end{eqnarray*}
where $ \langle \Pi \vert $ and {\bf P} are the momenta canonically conjugate to $ \vert \psi \rangle $ and {\bf R}, respectively:
\begin{equation}
\langle \Pi \vert 
\equiv
\frac{\delta L}{\delta\left(d\vert\psi\rangle/dt\right)}
=
i\frac{h}{2\pi}\langle\psi\vert,
\label{eq137}
\end{equation}
\begin{equation}
{\bf P}\equiv\frac{\partial L}{\partial\dot{{\bf R}}}=
M\dot{{\bf R}}.
\label{eq138}
\end{equation}
The time derivative of the analytical-mechanical Hamiltonian is given, 
with the help of Eqs.~(\ref{eq134}), (\ref{eq135}), (\ref{eq137}) and (\ref{eq138}), by
\begin{equation*}
\frac{dH}{dt}=\frac{{\bf P}}{M}\cdot\dot{{\bf P}}+
\dot{{\bf R}}\cdot\nabla_{{\mbox R}}
\langle\psi\vert V({\bf R}-\hat{{\bf r}})\vert\psi\rangle
=0,
\end{equation*}
which shows that the energy is conserved, as is expected. This also confirms that the quantum Hamiltonian $\hat{H}$ is the operator corresponding to energy.

Next we consider the conservation law of momentum. 
In discussing the conservation law for a quantum particle, we must investigate the expectation value of its momentum. The conservation law of momentum is therefore expressed as\begin{eqnarray*}
\frac{d}{dt}({\bf P}+\langle\hat{{\bf p}}\rangle)
&=&
-\nabla_{{\mbox R}}\langle V({\bf R}-\hat{{\bf r}})\rangle
+
\frac{2\pi}{ih}\left\langle\left[\hat{{\bf p}},\frac{h}{4\pi\theta m}
\hat{{\bf p}}^{2}+V({\bf R}-\hat{{\bf r}})\right]
\right\rangle 
\nonumber \\[8pt]
&=&
\left(\frac{2\pi\theta}{h}-1\right)\nabla_{{\mbox R}}
\langle V({\bf R}-\hat{{\bf r}})\rangle
\nonumber \\[8pt]
&=&
0,
\end{eqnarray*}
with the help of Eqs.~(\ref{eq135}),(\ref{eq125}), and (\ref{eqccrp}).
This equation shows that the translation generator is identical with the momentum operator and the total momentum is conserved, if and only if we put
\begin{equation}
\theta=\frac{h}{2\pi}.
\label{eq141}
\end{equation}
The constant $\theta$ must be a universal constant in order that momenta are additive. Therefore Eq.~(\ref{eq141}) shows again that Planck's constant must also be a universal constant.

From Eq.~(\ref{eq141}), the commutation relations (\ref{eqccr3}) between the 
position and momentum operators become the canonical commutation relations
\begin{equation}
\left[\hat{x}_{j},\hat{p}_{k}\right]=i\hbar\delta_{jk},
\label{eqccr4}
\end{equation}
as should be, where we have put
\begin{equation*}
\hbar \equiv \frac{h}{2\pi}.
\end{equation*}
Thus the full set of the canonical commutation relations (\ref{eqccr1}), (\ref{eqccr2}), and (\ref{eqccr4}) have derived from the principles of quantum mechanics. Furthermore the commutation relations between the components of angular momentum and between angular momentum and any vector operator $ \bf{\hat A} $ become 
\[
\left[\hat{L}_{j},\hat{L}_{k}\right]=i\hbar\varepsilon_{jkl}\hat{L}_{l},
\hspace{18pt}
\left[\hat{L}_{j},\hat{A}_{k}\right]
=
i\hbar\varepsilon_{jkl}\hat{A}_{l},
\]
from Eqs.~(\ref{eqCRLV}) and (\ref{eqCRLL}).

Thus substitution of Eq.~(\ref{eq141}) into Eq.~(\ref{eqHam2}) gives the final form of the Hamiltonian
\begin{equation*}
\hat{H}
=
\frac{\hat{{\bf p}}^{2}}{2m}
+
V ( {\bf{\hat r}}, {\bf{\hat p}} ),
\end{equation*}
which has the same form as that in classical mechanics. 
We have finally obtained the Schr\"odinger equation:
\begin{equation*}
i\hbar \frac{d}{dt} \vert \psi \rangle
=
\hat{H} \vert \psi \rangle
=
\left[ \frac{\hat{{\bf p}}^{2}}{2m} + V({\bf{\hat r}}, {\bf{\hat p}}) \right] \vert \psi \rangle.
\end{equation*}

It is crucially important to note that we have derived the Hamiltonian and the Schr\"odinger equation without being premised on classical mechanics, except for the definitions of dynamical variables.

\section{\label{sec9}CONCLUSION}

We have deductively established quantum mechanics on the basis of the six fundamental principles. We now summarize several essential points.

First, it is most important to note that the canonical commutation relations
\begin{equation*}
\left[\hat{x}_{j},\hat{p}_{k}\right]=i\hbar\delta_{jk},
\hspace{18pt}
\left[ {\hat x_j ,\hat x_k } \right] = 0,
\hspace{18pt}
\left[\hat{p}_{j},\hat{p}_{k}\right]=0
\end{equation*}
have been deduced from the principles.

We have defined momentum and angular momentum as the constants of motion that 
are derived from homogeneity and isotropy of space, respectively. 
The properties of momentum and angular momentum operators, which include the 
commutation relations between them, are necessarily established from this 
definition. It is therefore necessary that the canonical commutation relations hold.

Second, the Schr\"odinger equation that describes the time evolution of a state vector has necessarily been derived in the definitive form from the principles. 
Moreover, it has been verified that the operator that appears on the right side of the Schr\"odinger equation be the Hamiltonian or the operator corresponding to energy, 
since energy is the constant of motion that is derived from homogeneity of time. 
Furthermore, it has been shown that Planck's constant should appear on the 
left side of the equation by corresponding it to one of Einstein-de Broglie 
formulas.

We have shown that it is also necessary in the nonrelativistic theory that 
both of the Schr\"odinger equation and the canonical commutation relations 
include Planck's constant \footnote{This is trivial in the relativistic theory, but not in the nonrelativistic theory.}. Planck's constant has been introduced into the Schr\"odinger equation in accordance with one of Einstein-de Broglie formulas, 
$E=h\nu$, whereas Planck's constant in the canonical commutation relations has 
been introduced from the requirement that the translation generator is identical with momentum. 
It has also been derived both from the additivity of momenta and from that of Lagrangians that Planck's constant should be a universal constant.

Third, it have been shown that relations between dynamical variables in 
quantum mechanics have the same forms for those in classical mechanics such as 
the expressions for angular momentum 
$\hat{{\bf L}}=\hat{{\bf r}}\times\hat{{\bf p}}$, 
Hamiltonian $\hat{H}=\hat{{\bf p}}^{2}/2m + \hat V$, and the relation between momentum and velocity operators 
$\hat{{\bf p}}=m\hat{{\bf v}}$. 
It is important to note that we have derived them independently of classical 
mechanics. It is usually assumed that relations between dynamical variables in 
classical mechanics are also preserved in quantum mechanics. This, however, is not self-evident. For example, the momentum operator has no direct relation to the velocity operator in relativistic quantum mechanics. 

As a consequence, the process of construction of quantum mechanics on the 
basis of the fundamental principles shows that quantum mechanics is the subtle and 
inevitable theory. For example, quantum mechanics does not permit any other 
commutation relation except for the canonical commutation relations. The natural law is not selected arbitrarily but determined necessarily.

There are several quantization methods such as canonical, path integral, and random process. 
All of them are premised on the classical theory and are methods for deriving the quantum theory from the classical theory by quantizing it.  
The present formalism is not premised on the classical theory but is based on the fundamental principles that are independent of the classical theory. 
The rigorous quantum theory must not be premised on the classical theory that is approximate to the quantum theory.

\appendix*

\section{The condition that the equation of continuity is derived from the equation of motion }

We now find the condition such that the right side of Eq.~(\ref{eq100}) 
can be expressed as the divergence of a vector. 
The terms that do not include time derivatives on the right side of 
Eq.~(\ref{eq100}) can be rewritten as

\[
\begin{array}{l}
-\langle\psi\vert{\bf r}\rangle\langle{\bf r}\vert\hat{C}_{0}
\vert\psi\rangle+
\langle\psi\vert\hat{C}_{0}\vert{\bf r}\rangle\langle{\bf r}
\vert\psi\rangle
\\[8pt]
=
-\langle\psi\vert{\bf r}\rangle\langle{\bf r}\vert
\left[c_{00}+c_{02}\hat{{\bf p}}^{2}+c_{04}
\left(\hat{{\bf p}}^{2}\right)^{2}+\cdots\right]\vert\psi\rangle
\\[8pt]
\hspace{30pt}
+\langle\psi\vert\left[c_{00}+c_{02}\hat{{\bf p}}^{2}+c_{04}
\left(\hat{{\bf p}}^{2}\right)^{2}+\cdots\right]
\vert{\bf r}\rangle\langle{\bf r}\vert\psi\rangle
\\[8pt]
=
-\langle\psi\vert{\bf r}\rangle\left[-c_{02}\theta^{2}
\left\{\nabla^{2}\langle{\bf r}\vert\right\}+c_{04}\theta^{4}
\left\{\left(\nabla^{2}\right)^{2}\langle{\bf r}\vert\right\}+\cdots\right]
\vert\psi\rangle
\\[8pt]
\hspace{30pt}+\langle\psi\vert\left[-c_{02}\theta^{2}
\left\{\nabla^{2}\vert{\bf r}\rangle\right\}+c_{04}\theta^{4}
\left\{\left(\nabla^{2}\right)^{2}\vert{\bf r}\rangle\right\}+\cdots\right]
\langle{\bf r}\vert\psi\rangle
\\[8pt]
=
c_{02}\theta^{2}\left[\psi^{\ast}({\bf r})
\left\{\nabla^{2}\psi({\bf r})\right\}
-
\left\{\nabla^{2}\psi^{\ast}({\bf r})\right\}
\psi({\bf r})\right]
\\[8pt]
\hspace{30pt}
-
c_{04}\theta^{4}\left[\psi^{\ast}({\bf r})
\left\{\left(\nabla^{2}\right)^{2}\psi({\bf r})\right\}
-
\left\{\left(\nabla^{2}\right)^{2}\psi^{\ast}({\bf r})\right\}
\psi({\bf r})\right]
-
\cdots,
\end{array}
\]
where use has been made of Eqs.~(\ref{defCn}), (\ref{eqPosR}), and (\ref{eq34a}). 

It can easily be seen that the equation
\[
\psi^{\ast}\left\{(\nabla^2)^n\psi\right\}
-
\left\{(\nabla^2)^n\psi^{\ast}\right\}\psi=\nabla\cdot
{\bf J}_{2n}
\]
is generally valid, where
\[
\begin{array}{l}
{\bf{J}}_{2n}
= \left[ {\psi ^* \left\{ {\nabla \left( {\nabla ^2 } \right)^{n - 1} \psi } \right\} - \left\{ {\nabla \left( {\nabla ^2 } \right)^{n - 1} \psi ^* } \right\}\psi } \right] \\[8pt] 
\hspace{50pt}
- \left[ {\left( {\nabla \psi ^* } \right)\left\{ {\left( {\nabla ^2 } \right)^{n - 1} \psi } \right\} - \left\{ {\left( {\nabla ^2 } \right)^{n - 1} \psi ^* } \right\}\left( {\nabla \psi } \right)} \right] \\[8pt] 
\hspace{50pt}
+ \left[ {\left( {\nabla ^2 \psi ^* } \right)\left\{ {\nabla \left( {\nabla ^2 } \right)^{n - 2} \psi } \right\} - \left\{ {\nabla \left( {\nabla ^2 } \right)^{n - 2} \psi ^* } \right\}\left( {\nabla ^2 \psi } \right)} \right] \\[8pt] 
\hspace{50pt}
-  \cdots  \\[8pt] 
\hspace{50pt}
- \left[ {\left\{ {\nabla \left( {\nabla ^2 } \right)^{n/2 - 1} \psi ^* } \right\}\left\{ {\left( {\nabla ^2 } \right)^{n/2} \psi } \right\} - \left\{ {\left( {\nabla ^2 } \right)^{n/2} \psi ^* } \right\}\left\{ {\nabla \left( {\nabla ^2 } \right)^{n/2 - 1} \psi } \right\}} \right]
\end{array}
\]
for even $n$, and
\[
\begin{array}{l}
{\bf{J}}_{2n}  = \left[ {\psi ^* \left\{ {\nabla \left( {\nabla ^2 } \right)^{n-1} \psi } \right\} - \left\{ {\nabla \left( {\nabla ^2 } \right)^{n-1} \psi ^* } \right\} \psi } \right] \\[8pt]
\hspace{50pt}
- \left[ {\left( {\nabla \psi ^* } \right) \left\{ {\nabla \left( {\nabla ^2 } \right)^{n-1} \psi } \right\} - \left\{ {\nabla \left( {\nabla ^2 } \right)^{n-1} \psi ^* } \right\} \left( {\nabla \psi } \right)} \right] \\[8pt]
\hspace{50pt}
+ \left[ {\left( {\nabla ^2 \psi ^* } \right) \left\{ {\nabla \left( {\nabla ^2 } \right)^{n-2} \psi } \right\} - \left\{ {\nabla \left( {\nabla ^2 } \right)^{n-2} \psi ^* } \right\} \left( {\nabla ^2 \psi } \right)} \right] \\[8pt]
\hspace{50pt}
-  \cdots  \\[8pt]
\hspace{50pt}
+ \left[ \left\{ \left( {\nabla ^2 } \right)^{\frac{n-1}{2}} \psi ^*  \right\} \left\{ {\nabla \left( {\nabla ^2 } \right)^{\frac{n-1}{2}} \psi } \right\} - \left\{ {\nabla \left( {\nabla ^2 } \right)^{\frac{n-1}{2}} \psi ^* } \right\} \left\{ \left( {\nabla ^2 } \right)^{\frac{n-1}{2}} \psi  \right\} \right]
\end{array}
\]
for odd $n$.

Thus we obtain
\[
 - \left\langle {\psi }
 \mathrel{\left | {\vphantom {\psi  {\bf{r}}}}
 \right. \kern-\nulldelimiterspace}
 {{\bf{r}}} \right\rangle \left\langle {\bf{r}} \right|\hat C_0 \left| \psi  \right\rangle  + \left\langle \psi  \right|\hat C_0 \left| {\bf{r}} \right\rangle \left\langle {{\bf{r}}}
 \mathrel{\left | {\vphantom {{\bf{r}} \psi }}
 \right. \kern-\nulldelimiterspace}
 {\psi } \right\rangle  = \nabla  \cdot \left[ {c_{02} \theta ^2 {\bf{J}}_2  - c_{04} \theta ^4 {\bf{J}}_4  +  \cdots } \right],
\]
which can be expressed as the divergence of a vector.

The terms with a first-order derivative with respect to time on the right side of 
Eq.~(\ref{eq100}) are explicitly expressed, with the help of Eqs.~(\ref{defC'1}), (\ref{eqPosR}), and (\ref{eq34a}), as
\[
\begin{array}{l}
-i\langle\psi\vert{\bf r}\rangle\langle{\bf r}\vert
\hat{C}_{1}^{\prime}\displaystyle
\frac{d\vert\psi\rangle}{dt}-
\displaystyle
i\frac{d\langle\psi\vert}{dt}\hat{C}^{\prime}_{1}\vert{\bf r}\rangle
\langle{\bf r}\vert\psi\rangle
\\[8pt]
=
-i\langle\psi\vert{\bf r}\rangle\left[-c_{12}\theta^{2}
\left\{\nabla^{2}\langle{\bf r}\vert\right\}+c_{14}\theta^{4}
\left\{\nabla^{4}\langle{\bf r}\vert\right\}+\cdots\right]
\displaystyle
\frac{d\vert\psi\rangle}{dt}
\\[6pt]
\hspace{28pt}
\displaystyle
-i\frac{d\langle\psi\vert}{dt}
\left[-c_{12}\theta^{2}
\left\{\nabla^{2}\vert{\bf r}\rangle\right\}+c_{14}\theta^{4}
\left\{\nabla^{4}\vert{\bf r}\rangle\right\}+\cdots\right]
\langle{\bf r}\vert\psi\rangle
\\[8pt]
=ic_{12}\theta^{2}\left[\psi^{\ast}\left\{\nabla^{2}
\displaystyle
\frac{\partial\psi}{\partial t}\right\}+\left\{\nabla^{2}
\displaystyle
\frac{\partial\psi^{\ast}}{\partial t}\right\}\psi\right]
+\cdots,
\end{array}
\]
which cannot be expressed as the divergence of a vector. Therefore
\begin{equation*}
\hat{C}^{\prime}_{1}=0,
\end{equation*}
in order that Eq.~(\ref{eq100}) reduces to the equation of continuity. 

In the similar way, 
\begin{equation*}
\hat{C}_{n}=0,\hspace{40pt}n\geq 2.
\end{equation*}

\end{document}